\documentclass[a4paper,preprintnumbers,showpacs,onecolumn,superscriptaddress,nofootinbib,amsmath,amssymb,notitlepage]{revtex4-2}

\usepackage{mathtools,leftindex,tensor,mhchem}
\usepackage{booktabs}
\usepackage{siunitx}
\usepackage{enumitem}
\usepackage{times}
\usepackage{dcolumn}
\usepackage{mathrsfs}
\usepackage{comment}
\usepackage{hyperref}
\usepackage{amsmath,accents}
\usepackage{mathtools}
\usepackage{bbm}
\usepackage{bm}
\usepackage{amsfonts}
\usepackage{amssymb}
\usepackage{mathrsfs}
\usepackage[scr=rsfs,cal=boondox]{mathalfa}
\usepackage{wasysym}
\usepackage{graphicx}
\usepackage[]{subfigure}
\usepackage{filecontents}
\usepackage[dvipsnames]{xcolor}
\usepackage{bbold}
\usepackage[scr=rsfs,cal=boondox]{mathalfa}

\usepackage{stackengine}
\stackMath
\newcommand\tsup[2][2]{%
 \def\useanchorwidth{T}%
  \ifnum#1>1%
    \stackon[-1.3ex]{\tsup[\numexpr#1-1\relax]{#2}}{\mathchar"307E}%
  \else%
    \stackon[-1ex]{#2}{\mathchar"307E}%
  \fi%
}
\hypersetup{
    bookmarks=true,         
    pdftoolbar=true,        
    pdfmenubar=true,        
    pdffitwindow=true,     
    pdfstartview={FitH},     
    pdftitle={My title},     
    pdfauthor={author},      
    pdfsubject={Subject},    
    pdfcreator={Creator},    
    pdfproducer={Producer},  
    pdfkeywords={keyword1} 
    pdfnewwindow=true,       
    colorlinks=true ,        
    linkcolor=Blue  ,        
    citecolor=Blue,      
    filecolor=Blue,       
    urlcolor=Blue            
}

\newcommand{\arccosh}{\operatorname{arccosh}}

\newcommand{\ed}{\mathrm{d}}

\newcommand{\tQ}{\tilde{Q}}

\newcommand{\oalpha}[1]{\accentset{\circ}{\alpha}}
\newcommand{\obf}[1]{\accentset{\circ}{\mathbf{f}}}
\newcommand{\boR}[1]{\accentset{\circ}{\mathbf{R}}}
\newcommand{\obF}[1]{\accentset{\circ}{\mathbf{F}}}
\newcommand{\obPi}[1]{\accentset{\circ}{\mathbf{\Pi}}}

\usepackage{scalerel}
\usepackage{tikz}
\usetikzlibrary{svg.path}

\definecolor{orcidlogocol}{HTML}{A6CE39}
\tikzset{
  orcidlogo/.pic={
    \fill[orcidlogocol] svg{M256,128c0,70.7-57.3,128-128,128C57.3,256,0,198.7,0,128C0,57.3,57.3,0,128,0C198.7,0,256,57.3,256,128z};
    \fill[white] svg{M86.3,186.2H70.9V79.1h15.4v48.4V186.2z}
                 svg{M108.9,79.1h41.6c39.6,0,57,28.3,57,53.6c0,27.5-21.5,53.6-56.8,53.6h-41.8V79.1z M124.3,172.4h24.5c34.9,0,42.9-26.5,42.9-39.7c0-21.5-13.7-39.7-43.7-39.7h-23.7V172.4z}
                 svg{M88.7,56.8c0,5.5-4.5,10.1-10.1,10.1c-5.6,0-10.1-4.6-10.1-10.1c0-5.6,4.5-10.1,10.1-10.1C84.2,46.7,88.7,51.3,88.7,56.8z};
  }
}

\newcommand\orcidicon[1]{\href{https://orcid.org/#1}{\mbox{\scalerel*{
\begin{tikzpicture}[yscale=-1,transform shape]
\pic{orcidlogo};
\end{tikzpicture}
}{|}}}}

\begin{document}


\title{Shadow of a nonlinear electromagnetic generalized  Kerr-Newman-AdS black hole
}

\author{Mohsen Fathi\orcidicon{0000-0002-1602-0722}}
\email{mohsen.fathi@ucentral.cl}
\affiliation{Centro de Investigaci\'{o}n en Ciencias del Espacio y F\'{i}sica Te\'{o}rica, Universidad Central de Chile, La Serena 1710164, Chile\\
}


\begin{abstract}

In this work, we investigate the shadow properties of the Kerr-Newman-Anti-de Sitter black hole coupled to a nonlinear electrodynamics. We construct the shadow by employing the celestial coordinate approach for observers located at finite distances, accounting for the non-asymptotically flat nature of the spacetime. The size, distortion, area, and oblateness of the shadow are analyzed in terms of the black hole parameters, including the spin, effective charge, and nonlinearity parameter. We further explore observational constraints using the Event Horizon Telescope results for M87* and Sgr~A*. The angular diameters inferred from the observed images allow us to bound the parameter space of the model, showing that both spin and nonlinearity play key roles in shaping the shadow and in restricting the effective charge. Additionally, we examine the energy emission rate derived from the shadow radius and Hawking temperature, highlighting the influence of rotation and nonlinear electrodynamics on black hole evaporation. Our results demonstrate that the black hole model provides a physically consistent and observationally viable extension of the standard Kerr paradigm, with potential implications for near-horizon quantum processes and tests of gravity beyond general relativity.

\bigskip

{\noindent{\textit{keywords}}: Stationary black holes, Anti-de Sitter spacetime,  nonlinear electrodynamics, shadow 
}\\

\end{abstract}

\maketitle

\tableofcontents

\section{Introduction}\label{sec:intro}

The detection of gravitational waves from compact binary mergers by the LIGO \cite{LIGOScientific:2016aoc} and Virgo \cite{abbott_gwtc-3:_2023} collaborations has opened a new observational window into the strong-field regime of gravity. These events confirm the existence of massive compact remnants, which are often rotating. Consequently, within general relativity, stationary black hole solutions such as Kerr and Kerr-Newman (KN) play a central role, as they capture essential features of astrophysical compact objects.

In addition to rotation, electromagnetic effects can be important in the vicinity of highly magnetized objects such as neutron stars or magnetars. When the electromagnetic fields become extremely strong, Maxwell’s linear electrodynamics may no longer provide an adequate description, and nonlinear electrodynamics (NLE) models must be considered. These models, characterized by Lagrangians nonlinear in the electromagnetic invariants, have been widely employed to construct exact Einstein-NLE solutions. Such spacetimes can serve as models of strongly magnetized black holes, as testbeds for numerical simulations, and, in some cases, as regular black holes that avoid curvature singularities \cite{ayon-beato_regular_1998}. While static solutions with NLE sources have been extensively studied, obtaining fully rotating counterparts has proven to be significantly more challenging. Recent progress, however, includes a (KN) generalization with Euler-Heisenberg-type NLE corrections \cite{breton_rotating_2022}.

Broadly, NLE theories can be formulated as functions of the two Lorentz invariants $F$ and $G$. The Pleba\'nski classification \cite{plebanski_electromagnetic_1960}, further developed in Ref. \cite{boillat_nonlinear_1970}, has provided a systematic framework for such theories. Among their many implications is the modification of photon propagation: in NLE backgrounds, light rays follow null geodesics of effective ``optical metrics,'' leading in general to birefringence effects \cite{novello_geometrical_2000, obukhov_fresnel_2002, schellstede_causality_2016}.

The inclusion of a cosmological constant $\Lambda$ adds further relevance. A positive $\Lambda$ accounts for the accelerated expansion of the universe, while $\Lambda<0$ defines Anti-de Sitter (AdS) spacetimes that play a central role in the AdS/CFT correspondence \cite{witten_anti_1998}. In this context, black holes in AdS backgrounds acquire a dual description in terms of thermal field theories, motivating the study of AdS black holes with NLE sources.

Recently, exact stationary solutions that generalize the KN solution in the presence of NLE fields and a cosmological constant have been developed \cite{Garcia-Diaz:2021bao,garcia-diaz_adsds_2022,galindo-uriarte_nonlinear_2024}. These solutions describe rotating black holes endowed with mass, angular momentum, cosmological constant, effective electric charge, and a NLE parameter. The electromagnetic field tensors satisfy generalized Maxwell-like equations, while the Einstein equations are sourced by the NLE energy-momentum tensor. In particular, Ref.~\cite{galindo-uriarte_nonlinear_2024} constructed a family of Kerr-like geometries in which the NLE Lagrangian is determined consistently from the field invariants. This construction relies on aligning elements of the tetrad basis with the eigenvectors of the electromagnetic field tensor, supplemented by integrability conditions that constrain the electromagnetic potentials. The result is a broad class of exact rotating solutions with NLE corrections in AdS backgrounds.

In this work, we focus on one such solution (i.e., the KNAdS black hole with NLE. After briefly reviewing the key aspects of the construction based on Ref. \cite{galindo-uriarte_nonlinear_2024}, we analyze its horizon structure, ergoregion, and static limit (Sect.~\ref{sec:overview}). We then turn to photon dynamics (Sect.~\ref{sec:photon}), deriving the critical impact parameters of spherical photon orbits and constructing the corresponding shadows in non-asymptotically flat spacetimes. The shadow observables are subsequently employed (Sect.~\ref{sec:observables}) to constrain the BH parameters using recent Event Horizon Telescope (EHT) observations of M87* \cite{EventHorizonTelescope:2019pgp} and Sgr A* \cite{EventHorizonTelescope:2022wkp}. Finally, we discuss the implications of the energy emission rate and Hawking radiation for this class of black holes in Sect. \ref{sec:EnergyEmission}. We conclude in Sect. \ref{sec:conclusion}. Throughout, we adopt natural units ($G=c=1$), the metric signature $(-,+,+,+)$, and primes to denote derivatives with respect to the radial coordinate.




\section{The NLE generalized KNA\lowercase{d}S black hole and its properties}\label{sec:overview}

In the Boyer-Lindquist coordinates, the Kerr-like metrics possessing a cosmological constant $\Lambda$, are presented as
\begin{multline}
\ed s^2 = -\left(\frac{\Delta_r-a^2\Delta_\theta\sin^2\theta }{\Xi^2\Sigma}\right)\ed t^2+\left.\frac{\Sigma}{\Delta_r}\right.\ed r^2+\left.\frac{\Sigma}{\Delta_\theta}\right.\ed\theta^2+2a\sin^2\theta\left[\frac{\Delta_r-(a^2+r^2)\Delta_\theta}{\Xi^2\Sigma}\right]\ed t\ed\phi\\+
\sin^2\theta\left[\frac{(a^2+r^2)^2\Delta_\theta-a^2\Delta_r\sin^2\theta}{\Xi^2\Sigma}\right]\ed\phi^2,
    \label{eq:Kerr-like}
\end{multline}
in which
\begin{subequations}
    \begin{align}
      &  \Sigma(r,\theta) = r^2+a^2\cos^2\theta,\label{eq:Sigma}\\
      &  \Delta_\theta(\theta) = 1+\frac{\Lambda}{3}a^2\cos^2\theta,\label{eq:Deltatheta}\\
      &  \Xi = 1+\frac{\Lambda}{3}a^2,\label{eq:Xi}\\
      & \Delta_r(r) \doteq K(r) - 2 M r + r^2 + a^2 -\frac{\Lambda}{3}r^2(r^2+a^2),\label{eq:Deltar} 
    \end{align}
    \label{eq:metric_functions}
\end{subequations}
where $M$ and $a$, are, respectively, the mass and the spin parameter of the black hole, and the function $K(r)$ is related to the black hole's charge. In this sense, for the case of the KNAdS black hole, we assign $K(r)=Q_e^2+Q_m^2$, with $Q_e$ and $Q_m$ being the electric and magnetic charges. However, in the case that the black hole is coupled with NLE effects, this function has to be then determined by the NLE-Einstein equations. Note that, the determinant of the metric $\eqref{eq:Kerr-like}$ is obtained as
\begin{equation}
g=-g_{rr} g_{\theta\theta}\left(g_{t\phi}^2-g_{tt} g_{\phi\phi}\right) = -\frac{\Sigma^2\sin^2\theta}{\Xi^4}.
    \label{eq:g}
\end{equation}

\subsection{The NLE equations}\label{subsec:NLE}

We consider the Lagrangian $\mathcal{L}(F,G)$ that depends on the electromagnetic invariant $F$ and $G$, defined as 
\begin{eqnarray}
    && F= \frac{1}{4}F_{\mu\nu}F^{\mu\nu} = \frac{1}{2}\left(B^2-E^2\right),\label{eq:F}\\
    && G = \frac{1}{4}F_{\mu\nu}{\star F^{\mu\nu}} = -\vec{\bm{E}}\cdot\vec{\bm{B}},\label{eq:G}
\end{eqnarray}
in which $F_{\mu\nu} = 2A_{[\nu;\mu]}$ is the field strength tensor given in terms of the four-dimensional vector potential $\vec{\bm{A}}$. In the above equations, the three-dimensional electric and magnetic field vectors have been noted by $\vec{\bm E}$ and $\vec{\bm B}$, and $\star F_{\mu\nu} =\frac{1}{2}\sqrt{-g}\,\epsilon_{\mu\nu\alpha\beta} F^{\alpha\beta}$ is the dual tensor of $\bm F$, where $g$ is the determinant of the metric and $\epsilon_{\mu\nu\alpha\beta}$ is the Levi-Civita symbol. 

Note that, for the spacetime given in Eq. \eqref{eq:Kerr-like}, the components of $F_{\mu\nu}$ obey the alignment conditions $F_{r\phi} = -a \sin^2\theta F_{rt}$ and $F_{\theta t} = -1/a(a^2+r^2)F_{\theta\phi}$. Hence, only $F_{rt}$ and $F_{\theta t}$ are independent. By considering the definition of the field strength tensor respecting the vector potential, the alignment conditions can be recast as $A_{\phi,r} + a\sin^2\theta A_{t,r}=0$ and $A_{\theta,\phi}+1/a(a^2+r^2)A_{t,\theta}=0$. Furthermore, the integrability of $A_{\phi}$, provides the differential equation \cite{galindo-uriarte_nonlinear_2024}
\begin{equation}
    A_{t,\theta r}+\frac{2r}{\Sigma} A_{t,\theta}-\frac{2 a^2\sin\theta\cos\theta}{\Sigma}A_{t,r}=0,
    \label{eq:diff_Aphi}
\end{equation}
which has the general solution
\begin{equation}
A_t = \frac{X(r)+Y(\theta)}{\Sigma},
    \label{eq:At}
\end{equation}
with $X(r)$ and $Y(\theta)$ being arbitrary functions. In this context, hence, the alignment equations yield
\begin{equation}
A_\phi = -a \sin^2\theta\frac{X(r)}{\Sigma}-\frac{(a^2+r^2)}{a}\frac{Y(\theta)}{\Sigma}.
    \label{eq:Aphi}
\end{equation}
Note that, based on what shown earlier in this section, the KN black hole corresponds to $X(r)=Q_e r$ and $Y(\theta) = Q_m a \cos\theta$.

In general, and from the minimal coupling of the Lagrangian $\mathcal{L}(F,G)$ with Einstein's general relativity, provides the constitutive equations 
\begin{equation}
  \bm P = \mathcal{L}_F \bm F + \mathcal{L}_G\left(\star \bm F\right),
    \label{eq:P}
\end{equation}
in which $\bm P$ is a new skew symmetric tensor, and we have notated $\mathcal{L}_X\equiv\partial\mathcal L/\partial X$. Hence, the electromagnetic field equations will be ${P^{\mu\nu}}_{;\nu} = 0$ and ${\star F^{\mu\nu}}_{;\nu}=0$. Note that, same as $\bm F$, the dual tensor $\star \bm P$ is the curl of a potential, namely $\star \vec{\bm P}$. Hence we have, $\star P_{\mu\nu} = 2\star P_{[\nu;\mu]}$. Accordingly, the nonvanishing components of $\star \bm P$ are $\star P_{\theta t}$ and $\star P_{rt}$, and the electromagnetic field equations reduce to $\left(\sqrt{-g}\,P^{\phi\theta}\right)_{,\theta}+\left(\sqrt{-g}\,P^{\phi r}\right)_{,r}=0$. Hence, one can recast the constitutive equations \eqref{eq:P} as \cite{galindo-uriarte_nonlinear_2024}
\begin{equation}
\begin{pmatrix}
    \frac{F_{\theta t}}{a\sin\theta} & -F_{rt}  \\\\
    a \sin\theta F_{rt} & F_{\theta t}
\end{pmatrix}
\begin{pmatrix}
    \mathcal{L}_F   \\\\
    \mathcal{L}_G 
\end{pmatrix}
=
\begin{pmatrix}
    \star P_{t,r}  \\\\
    - {\star{P_{t,\theta}}}
\end{pmatrix},
    \label{eq:P_1}
\end{equation}
which can be used to express the derivatives of the Lagrangian. In Ref. \cite{galindo-uriarte_nonlinear_2024}, this equation has been exploited together with the electromagnetic equations, leading to obtaining
\begin{eqnarray}
    && \star P_{\phi,t} = -a \sin^2\theta\,{\star P_{t,r}},\label{eq:phipir}\\
    && \star P_{\phi,\theta} = -\frac{a^2+r^2}{a}\,{\star P_{t,\theta}},\label{eq:Pphitheta}
\end{eqnarray}
which can be solved to provide
\begin{eqnarray}
    && \star P_t = \frac{A(r)+B(\theta)}{\Sigma},\label{eq:Pt}\\
    && \star P_\phi = -a\sin^2\theta\frac{A(r)}{\Sigma}-\left(\frac{a^2+r^2}{a}\right)\frac{B(\theta)}{\Sigma},
\end{eqnarray}
in which $A(r)$ and $B(\theta)$ are arbitrary functions. These results show that the alignment conditions for $\bm P$ in the Kerr-like spacetime \eqref{eq:Kerr-like}, are independent of the kind of electrodynamics. Note that, the KN spacetime corresponds to choosing $A(r)=Q_m r$ and $B(\theta) = -Q_e a \cos\theta$. 

It should be notes that, since the electromagnetic potentials $\vec{\bm A}$ and $\star\vec{\bm P}$ are not independent of each other (linked by the constitutive equations),  their temporal parts in Eqs. \eqref{eq:At} and \eqref{eq:Pt} can be brought together in a single equation. In fact, the closure condition of the Lagrangian, i.e. $\ed^2\mathcal{L}=0$, demands that $\mathcal{L}_{,\theta r}=\mathcal{L}_{,r\theta}$. Using the chain rule to obtain the derivatives of the Lagrangian, we get \cite{galindo-uriarte_nonlinear_2024}
\begin{eqnarray}
    && \mathcal{L}_{,r} = \frac{\Xi^2}{a \sin\theta}\Big[({\star P_t)_{,r}} F_{\theta t,r}+({\star P_t})_{,\theta}F_{rt,r}\Big],\label{eq:L,r}\\
    && \mathcal{L}_{,\theta} = \frac{\Xi^2}{a\sin\theta}\Bigl[
    -\cot\theta\, F_{\theta t}({\star P_t})_{,r} + F_{\theta t,\theta}({\star P_t})_{,r} + F_{rt,\theta}({\star P_t})_{,\theta}
    \Bigr].
\end{eqnarray}
Considering these results in the closure condition, yields
\begin{equation}
A_{t, rr}\left[\frac{({\star P_t})_{,\theta}}{\sin\theta}\right]_{,\theta}-({\star P_t})_{,rr}\left[\frac{A_{t,\theta}}{\sin\theta}\right]_{,\theta} = 0,
    \label{eq:key}
\end{equation}
which is known as the \textit{key equation}, as called in Ref. \cite{garcia-diaz_adsds_2022}. With the help of this equation, one can constrain the functions $X(r), Y(\theta), A(r)$ and $B(\theta)$. Hence, in the case that the Lagrangian's explicit dependence on $F$ and $G$ is not known, the key equation has to be still satisfied by the electromagnetic potentials in the spacetime geometry.

\subsection{The NLE generalization of the KNA\lowercase{d}S spacetime}\label{subsec:NLE_gen}

In fact, once the electromagnetic potentials $A_t$ and $\star P_t$ are written in terms of the functions $X(r)$, $Y(\theta)$, $A(r)$ and $B(\theta)$, as in Eqs.~\eqref{eq:At} and \eqref{eq:Pt}, one can, in principle, generate an infinite class of solutions by introducing additional terms in these arbitrary functions. To explore this possibility, we adopt the following ansatz \cite{galindo-uriarte_nonlinear_2024}:
\begin{eqnarray}
    && X(r) = Q_e r + \sum_{n=-5}^{30} C_n r^n, \label{eq:Xr}\\
    && Y(\theta) = Q_m a \cos\theta + \sum_{s=-5}^{30} D_s \cos^s\theta, \label{eq:Ytheta}\\
    && A(r) = Q_m r + \sum_{l=-5}^{30} H_l r^l, \label{eq:Ar}\\
    && B(\theta) = -Q_e a \cos\theta + \sum_{k=-5}^{30} G_k \cos^k\theta,
\end{eqnarray}
where the constants $C_n$, $D_s$, $H_l$, and $G_k$ are constrained by the key relation \eqref{eq:key}. Despite this apparent generality, it turns out that only two distinct cases produce nontrivial solutions: the \emph{cubic} potentials \cite{garcia-diaz_adsds_2022} and the \emph{quartic} potentials \cite{galindo-uriarte_nonlinear_2024}. Other choices simply reduce to configurations where $A_t \propto \star P_t$, which correspond to the standard Maxwell theory and therefore do not yield new physics.

For the cubic case, the NLE modification of the metric function $K(r)$ in Eq.~\eqref{eq:Deltar} takes the form \cite{galindo-uriarte_nonlinear_2024}
\begin{equation}
    K(r) = \tilde{Q}^2\left(1-\beta r^2\right)^2,
    \label{eq:Kcubic}
\end{equation}
where $\tQ^2 \equiv Q_e^2 + Q_m^2$ denotes the effective total charge of the black hole, and $\beta$ is the nonlinear parameter. Clearly, in the limit $\beta=0$, the expression reduces to the standard KNAdS spacetime. This result, originally obtained in Ref.~\cite{garcia-diaz_adsds_2022}, demonstrates that depending on the values of $\beta$ and $\Lambda$, the solution can asymptotically describe de Sitter, anti--de Sitter, or flat geometries (the latter previously reported in Ref.~\cite{Garcia-Diaz:2021bao}). The resulting spacetime generally admits two horizons, and for asymptotically dS geometries, an additional cosmological horizon also emerges.

\begin{figure}[htp]
    \centering
    \includegraphics[width=8cm]{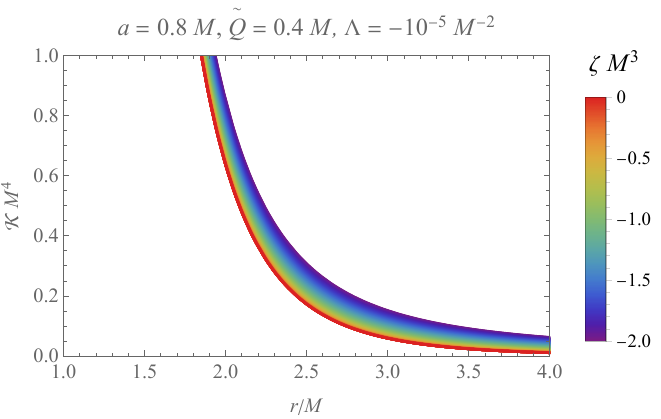} (a)\qquad
    \includegraphics[width=8cm]{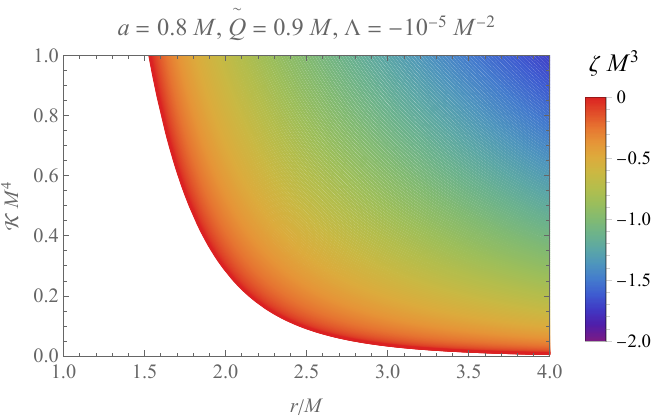} (b)
    \caption{Radial dependence of the Kretschmann scalar $\mathcal{K}$ for a rapidly rotating black hole, plotted for a range of negative values of the nonlinear parameter $\zeta$ and for two representative values of the effective charge $\tQ$.}
    \label{fig:Kretschmann}
\end{figure}

For the quartic potentials, Ref.~\cite{galindo-uriarte_nonlinear_2024} derived a different modification of $K(r)$:
\begin{equation}
    K(r) = \tQ^2\left(1+\zeta r^3\right),
    \label{eq:Kquartic}
\end{equation}
where $\zeta$ denotes the nonlinear parameter. In the system of units adopted here, one finds $\mathrm{dim}[\tQ] = \mathrm{length}$ and $\mathrm{dim}[\zeta] = \mathrm{length}^{-3}$. This structure defines a new family of KNAdS black holes generalized by NLE effects. For this case, the Kretschmann scalar, $\mathcal{K} = R_{\mu\nu\sigma\rho}R^{\mu\nu\sigma\rho}$, is presented in Appendix~\ref{app:B}, and its dependence on $\zeta$ for two different values of $\tQ$ is illustrated in Fig.~\ref{fig:Kretschmann}.
The plots reveal that $\mathcal{K}$ decays rapidly to zero at large distances from the black hole. Increasing the negative value of $\zeta$ enhances the curvature near the horizon, while larger $\tQ$ amplifies the overall scale of $\mathcal{K}$.

For the nonrotating case ($a=0$), the metric in Eq.~\eqref{eq:Kerr-like} reduces to a generalized Reissner--Nordström--AdS (RNAdS) black hole, which has been investigated for both $\Lambda=0$ and $\Lambda\neq 0$ in Refs.~\cite{filho_analysis_2025,araujo_filho_remarks_2025}. Some aspects of the NLE-modified KNAdS spacetime with $\Lambda=0$ were also studied in Ref.~\cite{galindo-uriarte_nonlinear_2024}, where it was shown that the introduction of the NLE parameter $\zeta$ can induce a third horizon resembling a cosmological one. In this case, the static limit corresponds to that of a NLE-corrected RN black hole with $\Lambda \neq 0$. Furthermore, analysis of the NLE energy-momentum tensor suggests that physically acceptable solutions require $\zeta < 0$ to satisfy the energy conditions.

In the following subsection, we continue our investigation of the full form of the solution provided in Ref. \cite{galindo-uriarte_nonlinear_2024}, called from now on, the KNAdS-NLE black hole, by considering the complete metric \eqref{eq:Kerr-like} with the functions given in Eq.~\eqref{eq:metric_functions}, restricted to $\Lambda < 0$, and adopting the quartic form of the structure function \eqref{eq:Kquartic}, appropriate to the AdS asymptotics.

\subsection{General structure of the black hole}\label{subsec:general_structure}

The horizons of the black hole are obtained by those hypersurfaces satisfying $\Delta_r=0$, which provides the quartic
\begin{equation}
r^4 + \sum_{i=0}^3 \mathrm{a}_i r^{i} = 0,
    \label{eq:Deltar=0}
\end{equation}
where
\begin{subequations}
\begin{align}
& \mathrm{a}_0 = -\frac{3}{\Lambda}\left(a^2+\tQ^2\right),\\
& \mathrm{a}_1 = \frac{6M}{\Lambda},\\
& \mathrm{a}_2 = a^2-\frac{3}{\Lambda},\\
& \mathrm{a}_3 = -\frac{3\tQ^2\zeta}{\Lambda}.
\end{align}
    \label{eq:a0123}
\end{subequations}
This equation has four roots $r_j,\,\, j=\overline{1,4}$ (see appendix \ref{app:A}). There is a possibility that all these roots are real and positive; in such condition, the black hole will possess four horizons; an inner horizon ($r_-$), and event horizon ($r_+$), and two cosmological horizons ($r_{1++}$ and $r_{2++}$), whose presence is the effect of the presence of the NLE parameter, $\zeta$, as the Schwarzschild-like AdS black holes without NLE, do not possess any cosmological horizons. In other cases, the roots of $\Delta_r=0$ may correspond to a fewer number of horizons.

\begin{figure}[h!]
    \centering
    \includegraphics[width=6cm]{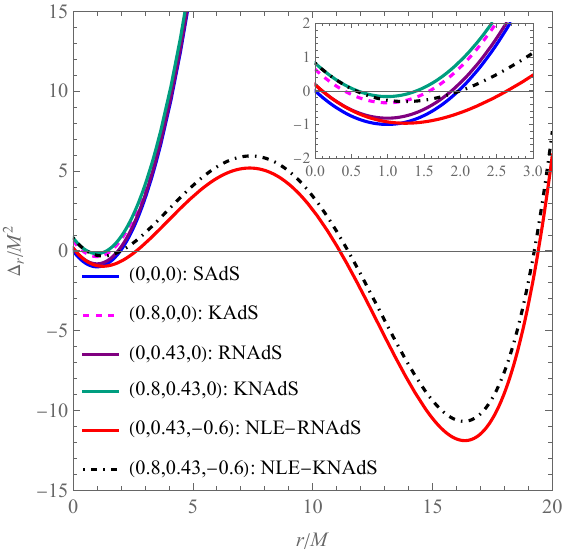} (a)\qquad
    \includegraphics[width=6cm]{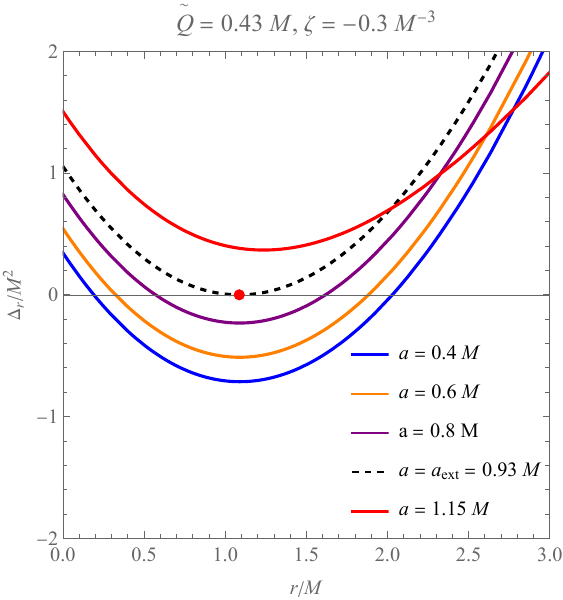} (b)
    \caption{(a) The radial profile of $\Delta_r$ presented for different categories regarding the coefficients $a, \tQ$ and $\zeta$. 
    Accordingly, each of the profiles correspond to a triple $(a/M, \tQ/M, \zeta M^3)$. (b) The radial profile of $\Delta_r$ for the NLE-KNAdS spacetime for different values for the spin parameter, by considering fixed values for $\tQ$ and $\zeta$. The red point indicates the location of the horizon $r_{\mathrm{ext}}$ of an EBH, at which $r_-=r_+$. The two cosmological horizons $r_{1,2++}$ haven not been shown in this panel as they are present in all cases. The red curve corresponds to a naked singularity with only the aforementioned two cosmological horizons.
    In both diagrams we have assumed $\Lambda=-10^{-2}\,M^{-2}$.}
    \label{fig:Deltar}
\end{figure}

\begin{figure}[htp]
    \centering
    \includegraphics[width=5.3cm]{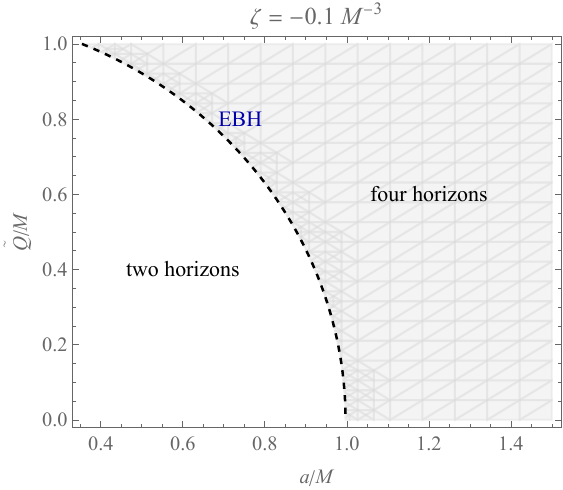} (a)\quad
    \includegraphics[width=5.3cm]{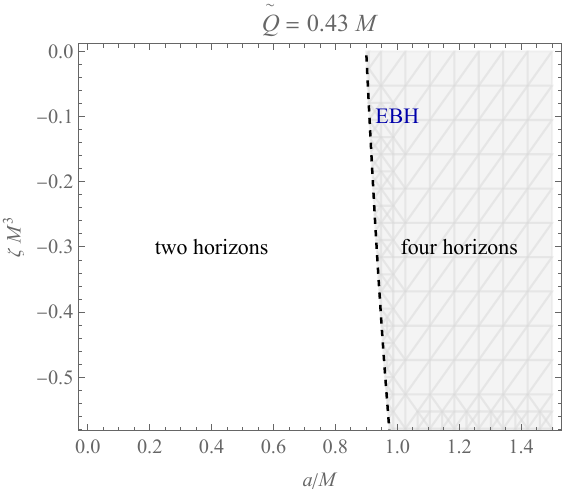} (b)\quad
    \includegraphics[width=5.3cm]{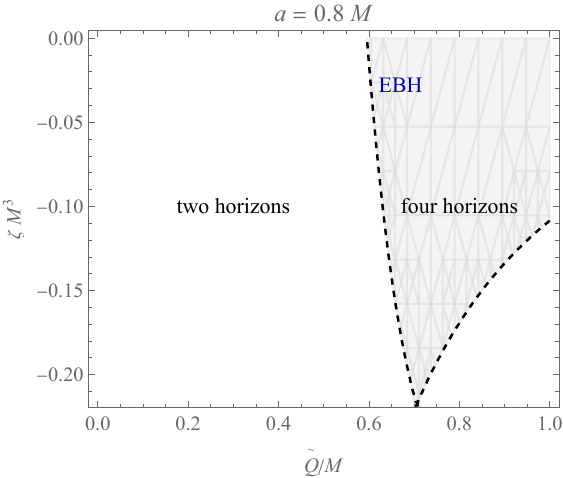} (c)
    \caption{Distinguishing between the regions with
$\mathrm{Disc}_{\Delta_r}>0$ (the shaded region) and those where $\mathrm{Disc}_{\Delta_r}<0$ (blank region),
    for the KNAdS-NLE spacetime, regarding changes in the spacetime parameters. 
   While the former proposes four horizons, the latter  offers only two horizons.
    The diagrams correspond to the mutual behaviors of the pairs (a) $(\tQ,a)$, (b) $(\zeta,a)$ and (c) $(\zeta,\tQ)$. 
    The dashed black curves correspond to the EBHs, for which, $\mathrm{Disc}_{\Delta_r}=0$, and the spacetime can possess three horizons.
    In all diagrams we have assumed $\Lambda=-10^{-2}\,M^{-2}$.}
    \label{fig:BH-NoBH}
\end{figure}

In order to elaborate on this, in Fig. \ref{fig:Deltar}(a), we have shown the radial profile of the metric function $\Delta_r$ for different values for its ingredients, each of which, presenting a different black hole model. 
As observed from the diagram, the cases corresponding to the Schwarzschild-AdS (SAdS), Kerr-AdS (KAdS), Reissner-Nordstr\"{o}m-AdS (RNAdS), and KNAdS black holes could possess only two horizons, whereas the RNAdS-NLE and KNAdS-NLE models could possess up to four horizons. It is then apparent that the cosmological horizons for the latter cases stem in the presence of the NLE ingredients in the spacetime geometry.  In Fig. \ref{fig:Deltar}(b), we have presented the radial profile of $\Delta_r$ for a NLE-KNAdS model when parameters $\tQ$ and $\zeta$ have been kept fixed. It is observed that by raise in the spin parameter, the black hole approaches the condition of extremality, in which, the first two horizons of the black hole, i.e. $r_\mp$, merge into a single horizon $r_\mathrm{ext}$ of an extremal black hole (EBH), while the cosmological horizons remain intact. To elaborate on this, we need to study the discriminant of the quartic $\Delta_r=0$, which is obtained as
\begin{multline}
\mathrm{Disc}_{\Delta_r} = \frac{1}{243}\Biggl\{
243 \zeta ^2 \tQ^4 \Biggl[4 \zeta  M \tQ^2 \left(-9 a^2+8 M^2-9 \tQ^2\right)-4 \left(a^2-M^2+\tQ^2\right)-27 \zeta ^2 \tQ^4 \left(a^2+\tQ^2\right)^2\Biggr]\\
-16 a^8 \Lambda ^5 \left(a^2+\tQ^2\right)-48 a^4 \Lambda ^4 \biggl[4 a^4+a^2 \left(M^2+12 \tQ^2\right)+8 \tQ^4\biggr]+324 \Lambda  \Biggl[4 \zeta  M \tQ^2 \left(-10 a^2+9 M^2-10 \tQ^2\right)\\
-4 \left(a^2-M^2+\tQ^2\right)+9 a^2 \zeta ^3 M \tQ^6 \left(a^2+\tQ^2\right)+\zeta ^2 \tQ^4 \biggl[-33 a^4+a^2 \left(4 M^2-69 \tQ^2\right)+6 \tQ^2 \left(M^2-6 \tQ^2\right)\biggr]\Biggr]\\
+108 \Lambda ^2 \Bigg[a^2 \zeta ^2 \tQ^4 \biggl[33 a^4+a^2 \left(M^2+69 \tQ^2\right)+36 \tQ^4\biggr]+4 \zeta  M \tQ^2 \biggl[44 a^4+a^2 \left(68 \tQ^2-9 M^2\right)+24 \tQ^4\biggr]\\
-4 \biggl[4 a^4+12 \tQ^2 \left(a^2-3 M^2\right)-33 a^2 M^2+27 M^4+8 \tQ^4\biggr]\Biggr]+36 \Lambda ^3 \Biggl[a^6 \zeta ^2 \tQ^4 \left(a^2+\tQ^2\right)-40 a^4 \zeta  M \tQ^2 \left(a^2+\tQ^2\right)\\
-4 \biggl[6 a^6+11 a^4 \left(3 M^2+2 \tQ^2\right)+4 a^2 \left(9 M^2 \tQ^2+8 \tQ^4\right)+16 \tQ^6\biggr]\Biggr]
\Biggr\}.
    \label{eq:discriminant}
\end{multline}
Accordingly, the extremality situation occurs when the condition $\mathrm{Disc}_{\Delta_r}=0$ holds, leading to a degenerate root for $\Delta_r=0$. At this point, the spin parameter gains its extremal value $a_\mathrm{ext}$. By increasing the spin parameter to values $a>a_\mathrm{ext}$, we have then $\mathrm{Disc}_{\Delta_r}<0$, and hence, the quartic $\Delta_r=0$ will possess two complex conjugate roots and two real roots. In such a condition, two of the horizons will vanish. In the case that these two horizons are the inner and the event horizons (as shown in Fig. \ref{fig:Deltar}(b)), a naked singularity would be generated possessing only two cosmological horizons $r_{1,2++}$. However, this is not always the case for the KNAdS-NLE spacetime (see below). In fact, the equation $\mathrm{Disc}_{\Delta_r}=0$ can be solved for $\tQ$ and $\zeta$, when the other spacetime parameters are kept fixed, in order to obtain their extremal values, $\tQ_{\mathrm{ext}}$ and $\zeta_\mathrm{ext}$. Based on this, one can distinguish the conditions under which only two horizons can form, based on these extremal values as the limits. This is what which has been shown in the diagrams in Fig. \ref{fig:BH-NoBH}, where we have shown how spacetimes with four horizons and those with only two horizons are separated, once the spacetime parameters are changed.
Also in Tables \ref{tab:1} and \ref{tab:2}, we have calculated the position of the horizons for different cases of the spacetime parameters.
\begin{table}
    \centering
    \begin{tabular}{c||c|c|c|c}
             $\zeta \,(M^{-3})$ & $r_-\, (M)$ & $r_+\, (M)$  & $r_{1,++}\, (M)$ & $r_{2++}\, (M)$ \\
         \hline\hline
         $-0.1$ & 0.325  & 1.719 & --  & -- \\
        $-0.3$ &  0.324 & 1.883 & --  &  -- \\
         $-0.45$&  0.323&  2.055& --  & -- \\
         $-0.584$&  0.323&  2.281& 14.901 & 14.901 \\
        $-0.62$ & 0.323&  2.362& 9.782 & 21.924\\
         $-0.798$& 0.322& 3.730 & 3.730 & 36.50\\
        $-0.9$ & 0.321 & -- & -- & 43.305\\
    \end{tabular}
    \caption{Te location of the horizons for the KNAdS-NLE black hole given $a = 0.6 M$, $\tQ = 0.43 M$ and $\Lambda=-10^{-2} M^{-2}$, for various assumed values for the $\zeta$-parameter. The cases of EBHs corresponding to black holes with three horizons, and those where the cosmological horizons disappear have been considered as well.}
    \label{tab:1}
\end{table}
\begin{table}
    \centering
    \begin{tabular}{c||c|c|c|c}
             $\tQ \,(M)$ & $r_-\, (M)$ & $r_+\, (M)$  & $r_{1,++}\, (M)$ & $r_{2++}\, (M)$ \\
         \hline\hline
         $0.1$ & 0.206  & 1.719 & --  & -- \\
        $0.2$ &  0.225 & 1.841 & --  &  -- \\
         $0.3$&  0.258&  1.946& --  & -- \\
         $0.424$&  0.319&  2.285& 14.90 & 14.90 \\
        $0.498$ & 0.368&  3.661& 3.661 & 36.966\\
         $0.6$& 0.453& -- & -- & 59.956\\
        $0.9$ & 0.644 & -- & -- & 143.739\\
    \end{tabular}
    \caption{Te location of the horizons for the KNAdS-NLE black hole given $a = 0.6 M$, $\zeta = - 0.6 M^{-3}$ and $\Lambda=-10^{-2} M^{-2}$, for various assumed values for the $\tQ$-parameter. 
    {The cases where $r_+$ and $r_{1++}$ would vanish, as well as the EBHs have been considered.}}
    \label{tab:2}
\end{table}
As inferred from the tables, there are cases where the spacetime can possess only two horizons, but still a black hole could continue to exist. In such cases (see Table \ref{tab:1}), the situation does not correspond to a naked singularity, instead it corresponds to a KNAdS-NLE black hole with vanishing cosmological horizons. In this context, EBHs could correspond to cases where the cosmological horizons would merge, or $r_+$ and $r_{1++}$ would merge, leaving a black hole with three horizons. Note that, by observing the data in both tables, one can notice that, there are no naked singularities present under such conditions for the spacetime parameters, even in the cases that they exceed their extremal values. Hence, the singularity is always censored by at least one horizon.

\begin{figure}[h!]
    \centering
    \includegraphics[width=5.3cm]{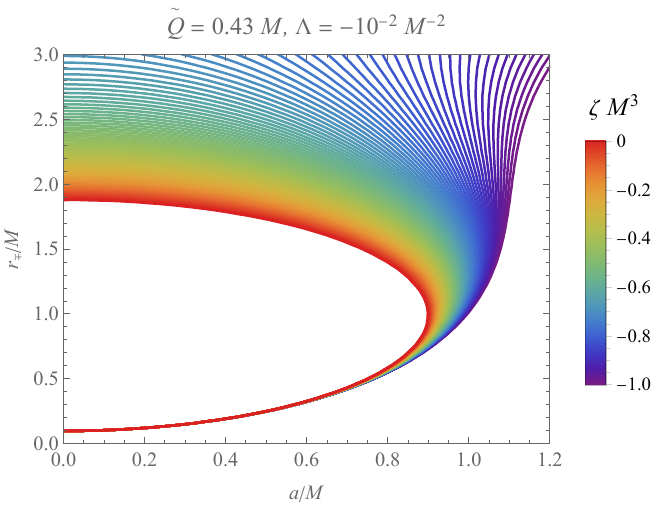} (a)\quad
    \includegraphics[width=5.3cm]{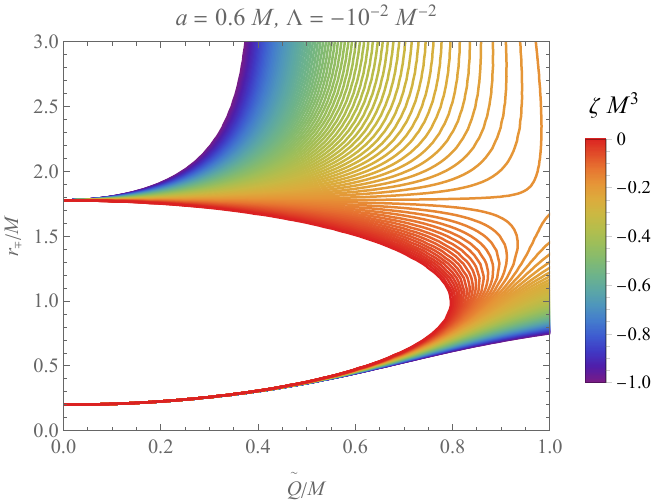} (b)\quad
    \includegraphics[width=5.3cm]{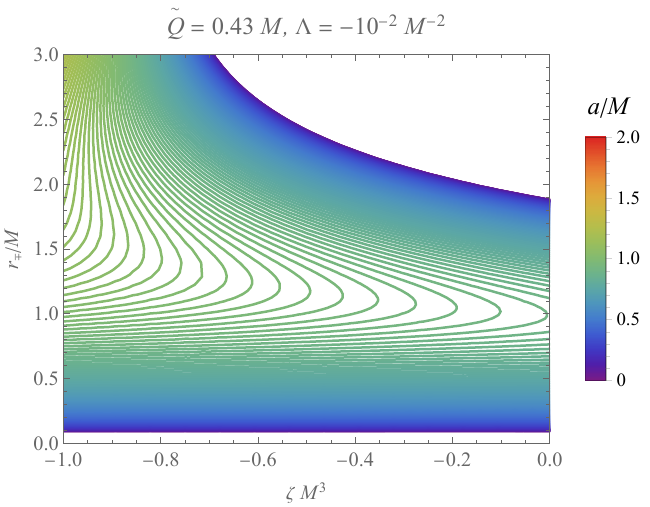} (c)
    \caption{The behavior of $r_\mp$, regarding the changes in the spacetime parameters.}
    \label{fig:Deltar=0}
\end{figure}

To further demonstrate the behavior of the black hole horizons, in Fig. \ref{fig:Deltar=0}, we have shown the profiles of the first two solutions of the quartic $\Delta_r=0$, i.e. $r_\mp$, for changes in the three spacetime parameters $a, \tQ$ and $\zeta$. 
As inferred from the diagrams, in non of the cases the singularity is exposed, and it is always concealed behind, at least, a horizon. Hence, the formation of the naked singularity is not always possible, and the conditions should be in particular in favor of its formation, such as those presented in Fig. \ref{fig:Deltar}(b).

\subsection{The static limit and the ergoregions}\label{subsec:ergo}

It is crucial to highlight that, beyond the event horizons, stationary black holes also possess specific hypersurfaces where static observers—those following worldlines given by $\bm{u} = (-g_{tt})^{-1/2} \bm{\chi}^t$—are not physically allowed. Here, $\bm{\chi}^t$ denotes the timelike Killing vector field associated with the time coordinate. These hypersurfaces, known as \textit{static limit surfaces}, mark the boundary where the observer's four-velocity $\bm{u}$ becomes null. 

The corresponding radial positions of these static limits, denoted by $r_{\rm{sl}}$, are determined by solving the condition $g_{tt} = 0$ using the line element from Eq.~\eqref{eq:Kerr-like}. Consequently, they can be found by replacing $a \rightarrow a\cos\theta$ in the roots of the equation $\Delta_r = 0$, namely $r_\mp$ and $r_{1,2++}$. In this context, we identify two sets of hypersurfaces with radii $r_{\rm{sl}_\mp}$ and $r_{\rm{sl}_{1,2++}}$, which define the permissible domain for the existence of static observers.

These surfaces follow the inequality$
r_{\rm{sl}_-} \leq r_- \leq r_+ \leq r_{\rm{sl}_+} \leq r_{\rm{sl}_{1++}} \leq r_{1++} \leq r_{\rm{sl}_{2++}} \leq r_{2++}$. Accordingly, the region outside the event horizon, where no static observers can reside, is bounded by $r_+ \leq r \leq r_{\rm{sl}_+}$ and is referred to as the \textit{outer ergoregion}. Similarly, the region $r_{\rm{sl}_-} \leq r \leq r_-$ defines the \textit{inner ergoregion}. Within these domains, the timelike Killing vector becomes spacelike, and static trajectories are not possible.

\begin{figure}[htp]
    \centering
    \includegraphics[width=3.8cm]{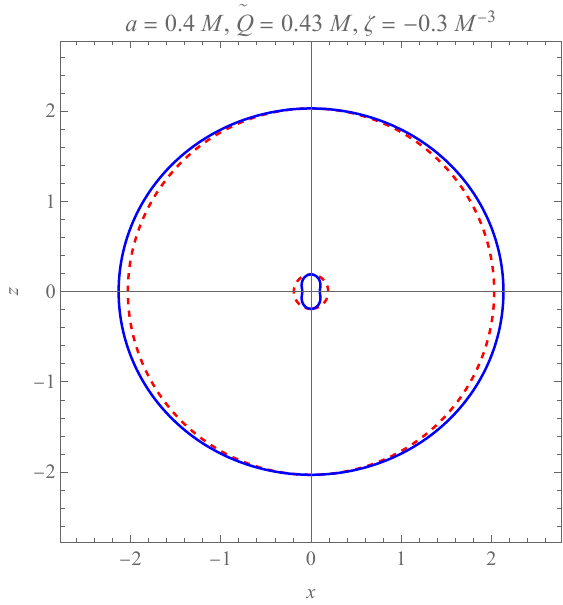} (a)\quad
    \includegraphics[width=3.8cm]{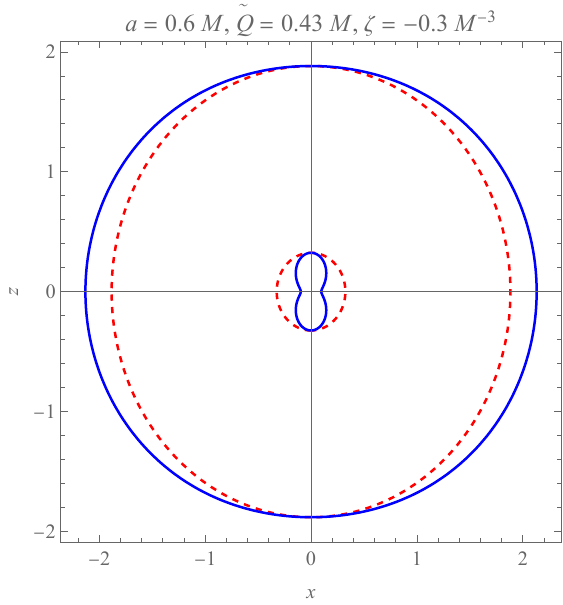} (b)\quad
    \includegraphics[width=3.8cm]{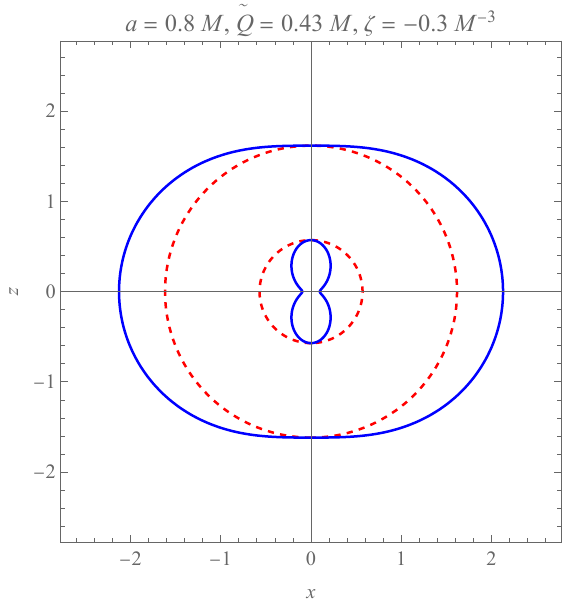} (c)\quad
    \includegraphics[width=3.8cm]{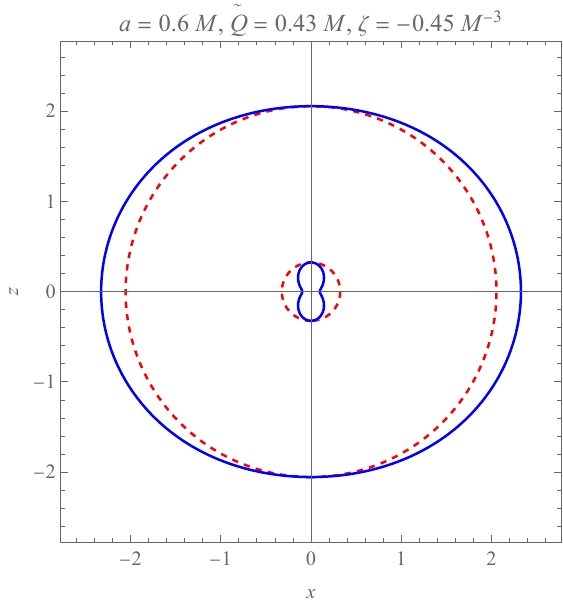} (d)
    \includegraphics[width=3.8cm]{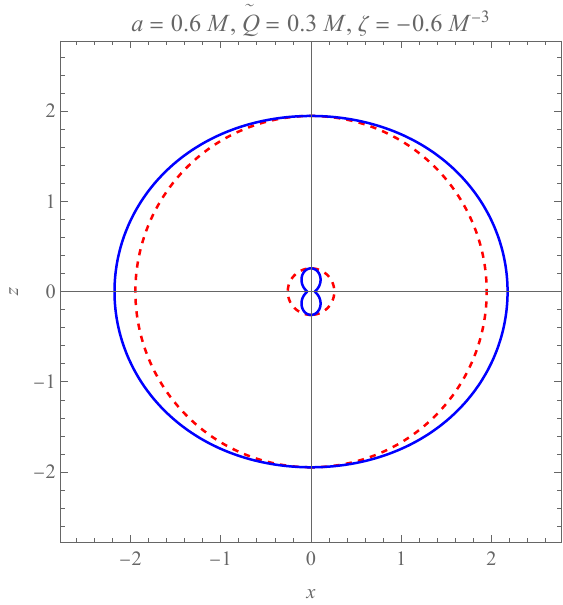} (e)\quad
    \includegraphics[width=3.8cm]{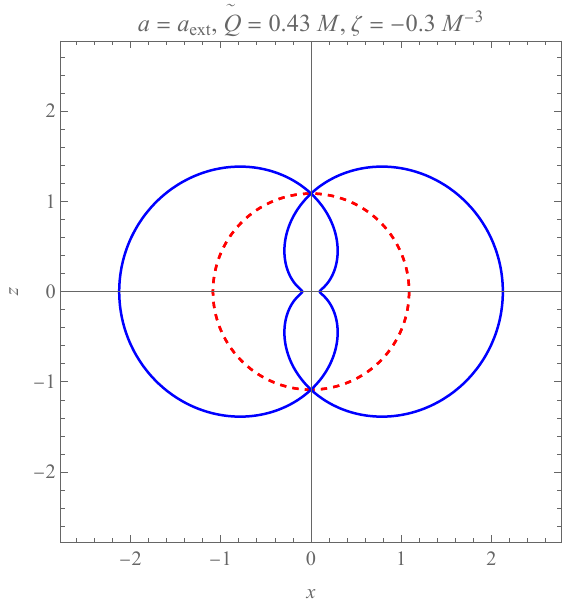} (f)\quad
    \includegraphics[width=3.8cm]{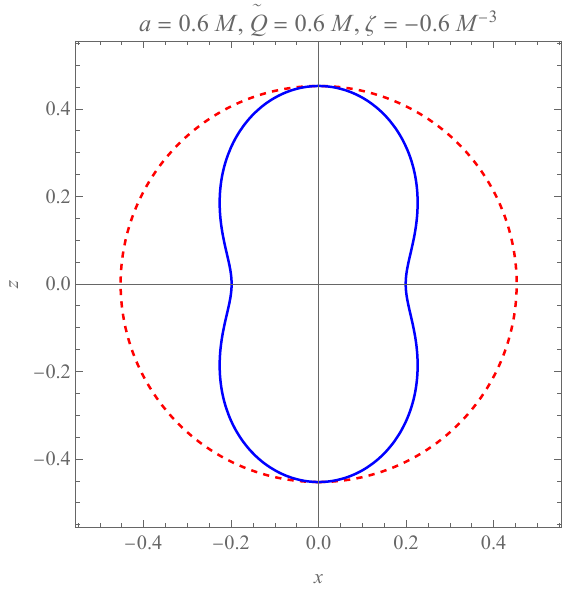} (g)\quad
    \includegraphics[width=3.8cm]{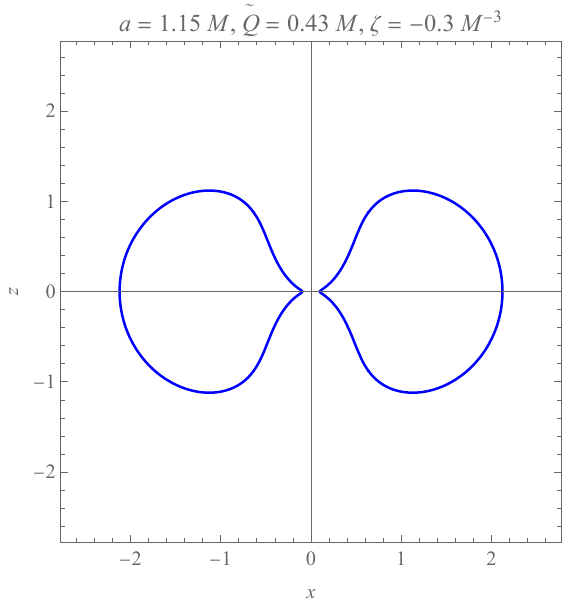} (h)
    \caption{The ergoregions for the KNAdS-NLE black hole, by assuming different values for the spacetime parameters, in accordance to the initial data considered in Fig. \ref{fig:Deltar}(b), and Tables \ref{tab:1} and \ref{tab:2}.
    The cases of EBH and the naked singularity have been shown, respectively, in panels (f) and (h).
    In all diagrams we have assumed $\Lambda=-10^{-2}\,M^{-2}$.}
    \label{fig:ergo}
\end{figure}

Using Cartesian coordinates defined as $x = r\sin\theta\cos\phi$, $y = r\sin\theta\sin\phi$, and $z = r\cos\theta$, Fig.~\ref{fig:ergo} illustrates examples of the ergoregions near a KNAdS-NLE black hole, plotted for different values of the spacetime parameters. For clarity in visualization, the ergoregions associated with the cosmological horizons are omitted.
The diagrams presented are based on parameter choices consistent with those used in Fig.~\ref{fig:Deltar}(b) as well as Tables~\ref{tab:1} and \ref{tab:2}. Consequently, configurations representing an EBH with coinciding horizons ($r_- = r_+$), and a naked singularity (where $r_\mp$ vanish), are also illustrated. As can be seen, increasing the spin parameter leads to an expansion of the ergoregion. This region is of particular interest due to the potential for energy extraction via the Penrose process~\cite{penrose_extraction_1971}.

\begin{figure}[h!]
    \centering
    \includegraphics[width=7cm]{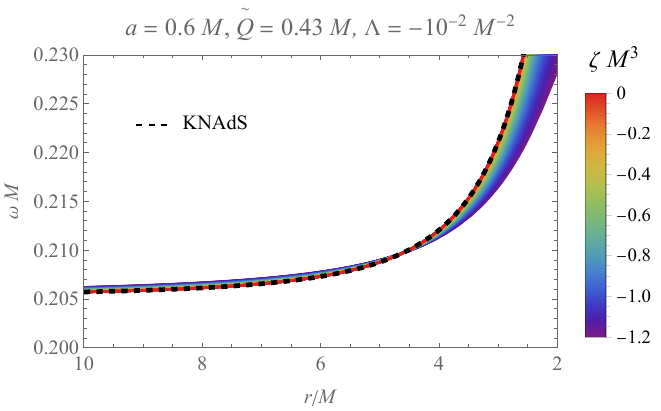}
    \caption{The radial profile of the angular velocity $\omega$ for some values of the black hole parameters, versus changes in the nonlinear parameter $\zeta$.}
    \label{fig:omega}
\end{figure}

It is also fruitful to note that, the frame-dragging of rotating black holes, becomes apparent from the off-diagonal terms existing in their spacetime metrics (such as the $g_{t\phi}$ term in the line element \eqref{eq:Kerr-like}). Due to this effect, the zero angular momentum observers (ZAMOs), would corotate with the black hole, by entering its ergoregion outside the event horizon. For the spacetime \eqref{eq:Kerr-like}, this corotation is performed by the angular velocity \cite{pugliese_observers_2018}
\begin{equation}
\omega = \frac{\ed\phi}{\ed t} = -\frac{g_{t\phi}}{g_{\phi\phi}} = \frac{a \Bigl[\left(a^2+r^2\right) \Delta_\theta-\Delta_r\Bigr]}{\left(a^2+r^2\right)^2 \Delta_\theta-a^2 \sin^2\theta\Delta_r}.
    \label{eq:omega}
\end{equation}
This quantity monotonously increases as the observer approaches the black
hole (see Fig. \ref{fig:omega}) 
and, ultimately at the event horizon, where the observer begins
corotation with the black hole, takes its maximum value, which is $\Omega_+ = \omega(r_+)$, which is obtained as
\begin{equation}
\Omega_{+} = \frac{a}{a^2+r_+^2}.
    \label{eq:Omega_KNAdS}
\end{equation}
This form is in common for all spacetimes which have Kerr-like rotation structure. Hence, $\Omega_+$ is the angular velocity of the black hole's event horizon, and is the value, at which, the ZAMOs would corotate by the black hole. In the case of the Kerr black hole, this value becomes $\Omega_+^{\mathrm{Kerr}} = a/2Mr_+$.



\section{Photon orbits in the NLE-KNA\lowercase{d}S spacetime and the black hole shadow}\label{sec:photon}

This section presents a detailed investigation into the behavior of the black hole shadow formed by null geodesics in the vicinity of the KNAdS-NLE black hole. The analysis is carried out using the standard Lagrangian approach, where the equations of motion are derived from the Hamilton-Jacobi formalism. These equations are then numerically integrated to determine the shadow structure.

The Hamilton-Jacobi equation for a particle in curved spacetime takes the form
\begin{equation} \label{e33}
    \frac{\partial S}{\partial \lambda} = -H,
\end{equation}
where $S$ denotes the Jacobi action, $\lambda$ is the affine parameter along the geodesics, and $x^{\alpha}$ are the spacetime coordinates. The Hamiltonian $H$ for geodesic motion in vacuum is given by
\begin{equation} \label{e34}
    H = \frac{1}{2} g^{\alpha \beta} \frac{\partial S}{\partial x^{\alpha}} \frac{\partial S}{\partial x^{\beta}}.
\end{equation}
Substituting this into Eq.~\eqref{e33}, the equation becomes
\begin{equation} \label{e35}
    \frac{\partial S}{\partial \lambda} = -\frac{1}{2} g^{\alpha \beta} \frac{\partial S}{\partial x^{\alpha}} \frac{\partial S}{\partial x^{\beta}}.
\end{equation}
Using the four-momentum definition $p_\alpha = \partial S / \partial x^\alpha$, and following Carter’s method of separability \cite{Carter:1968,Chandrasekhar:2002}, the action can be separated as
\begin{equation} \label{e36}
    S = \frac{1}{2} \mu^2 \lambda - E t + L \phi + S_r(r) + S_\theta(\theta),
\end{equation}
where $\mu$ is the rest mass of the test particles, and $\lambda$ is the affine parameter associated with the geodesics. For photons, we set $\mu = 0$. The symmetries of the spacetime imply conserved quantities: the energy $E$ and the angular momentum $L$, defined by
\begin{align}
    E &= -p_t = g_{tt} \dot{t} + g_{t\phi} \dot{\phi}, \\
    L &= p_\phi = g_{t\phi} \dot{t} + g_{\phi\phi} \dot{\phi}.
\end{align}
Applying the metric of the KNAdS-NLE black hole, the geodesic equations describing null trajectories are given by \cite{Slany:2020jhs, PhysRevD.81.044020,StuchlikEPJC2018}
\begin{align}
    \Sigma \frac{\ed t}{\ed\lambda} &= \frac{a \Xi^2}{\Delta_\theta} \left(L - a E \sin^2\theta\right) + \frac{\left(a^2 + r^2\right) \Xi^2}{\Delta_r} \Big[ (a^2 + r^2) E - a L \Big], \label{eq:tdot} \\
    \Sigma \frac{\ed r}{\ed\lambda} &= \epsilon_r \sqrt{R(r)}, \label{eq:rdot} \\
    \Sigma \frac{\ed\theta}{\ed\lambda} &= \epsilon_\theta \sqrt{\Theta(\theta)}, \label{eq:thetadot} \\
    \Sigma \frac{\ed\phi}{\ed\lambda} &= \frac{\Xi^2}{\Delta_\theta} \left(L \csc^2\theta - a E\right) + \frac{a \Xi^2}{\Delta_r} \Big[ (a^2 + r^2) E - a L \Big], \label{eq:phidot}
\end{align}
where $\epsilon_r = \epsilon_\theta = \pm 1$, and
\begin{subequations}
\begin{align}
    R(r) &= \Xi^2 \Bigl[ \Bigl( (a^2 + r^2) E - a L \Bigr)^2 - \Delta_r \Bigl( (L - a E)^2 + \mathcal{O} \Bigr) \Bigr], \label{eq:R} \\
    \Theta(\theta) &= \Bigl[ (L - a E)^2 + \mathcal{O} \Bigr] \Delta_\theta + \Xi^2 \cos^2\theta \left( a^2 E^2 - L^2 \csc^2\theta \right), \label{eq:Theta}
\end{align}
\end{subequations}
with $\mathcal{O} = \mathscr{Q} - (L - a E)^2$ representing the generalized Carter's constant \cite{Carter:1968rr}, and $\mathscr{Q}$ is the standard separation constant. The null trajectories of photons are thus fully described by Eqs.~\eqref{eq:tdot}–\eqref{eq:phidot}, with their behavior governed by the dimensionless impact parameters $\xi = L / E$ and $\eta = \mathscr{Q} / E^2$ \cite{Chandrasekhar:2002}. These trajectories can be categorized into three types: deflected (scattering) paths, spherical photon orbits, and plunging paths into the black hole.

Now rewriting the radial and angular potentials in terms of $\xi$ and $\eta$, we get
\begin{subequations}
\begin{align}
    R(r) &= E^2 \Xi^2 \Bigl[ \Bigl( (a^2 + r^2) - a \xi \Bigr)^2 - \Delta_r \Bigl( (\xi - a)^2 + \eta \Bigr) \Bigr], \label{eq:Rr} \\
    \Theta(\theta) &= E^2 \Bigl[ \eta \Delta_\theta + \Xi^2 \cos^2\theta \left( a^2 - \xi^2 \csc^2\theta \right) \Bigr]. \label{eq:Thetatheta}
\end{align}
\end{subequations}
The next subsections focus on analyzing the black hole shadow in the KNAdS-NLE geometry, using these impact parameters and geodesic equations to probe observable features.

\subsection{Orbits of constant radius and the critical impact parameters}\label{subsec:photonregions}

The investigation of black hole shadows is deeply rooted in the analysis of photon trajectories, particularly those maintaining a constant radial coordinate near the event horizon. These paths, termed \emph{spherical photon orbits}, are instrumental in delineating whether a photon escapes to infinity or is captured by the black hole. Due to their inherent instability, such orbits give rise to an infinite set of photon rings that define the apparent silhouette of the black hole.

In static spacetimes like the Schwarzschild geometry, these orbits lie entirely in a fixed plane. However, for rotating black holes, frame-dragging effects modify this structure, leading to a more intricate \emph{photon region} where spherical photon trajectories are no longer confined to a single plane. This region, delimited by the innermost and outermost circular photon orbits, encompasses the full domain of spherical photon motion. This behavior has been extensively explored within the Kerr spacetime framework (see Refs.~\cite{Chandrasekhar:2002,Bardeen:1972a,Bardeen:1973b}), and further investigations have expanded our understanding of such orbits in both Kerr and Kerr-like geometries (see Refs.~\cite{stoghianidis_polar_1987,cramer_using_1997,Teo:2003,Johannsen:2013,Grenzebach:2014,Perlick:2017,charbulak_spherical_2018,Johnson_universal_2020,Himwich:2020,Gelles:2021,Ayzenberg:2022,Das:2022,fathi_spherical_2023,ANJUM2023101195,Chen:2023,andaru_spherical_2023}).

Following the standard formalism, spherical photon orbits at a fixed radius $r_p$ are determined by the conditions \cite{Teo:2003}:
\begin{equation} \label{e39}
    R(r_p) = R'(r_p) = 0,
\end{equation}
where $R(r)$ denotes the radial potential. These constraints lead to the following critical impact parameters associated with spherical photon motion:
\begin{align}
    \xi_p &= \left( \frac{\Delta_r'(r^{2}+a^{2}) - 4\Delta_r r}{a \Delta_r' \Xi} \right)_{r_p}, \label{e46} \\
    \eta_p &= \left( \frac{16r^{2} \Delta_r (a^{2} - \Delta_r) - r^{4} (\Delta_r')^{2} + 8r^{3} \Delta_r \Delta_r'}{a^{2} (\Delta_r')^{2} \Xi^2} \right)_{r_p}, \label{e47}
\end{align}
where $\xi_p$ and $\eta_p$ denote the dimensionless impact parameters associated with the conserved angular momentum and the Carter constant, respectively.

When restricting to the equatorial plane, the spherical photon condition simplifies to $\Theta(\pi/2) = 0$, implying $\eta_p = 0$. For the KNAdS-NLE black hole, this condition yields the sixth-order polynomial:
\begin{equation}
    r^6 + \sum_{j=0}^{5} \mathrm{b}_j r^j = 0,
    \label{eq:rp_eq}
\end{equation}
with the coefficients given by:
\begin{subequations}
    \begin{align}
        \mathrm{b}_0 &= \frac{16 \left(a^2 + \tQ^2\right)}{\zeta^2 \tQ^2}, \\
        \mathrm{b}_1 &= -\frac{16 M \left(a^2 + 3\tQ^2\right)}{\zeta^2 \tQ^4}, \\
        \mathrm{b}_2 &= \frac{4 \left(27 M^2 - 4\tQ^2 \left(a^2 \Lambda - 3\right)\right)}{3 \zeta^2 \tQ^4}, \\
        \mathrm{b}_3 &= \frac{8 \left(M \left(a^2 \Lambda - 3\right) + \zeta \tQ^2 \left(\tQ^2 - a^2\right)\right)}{\zeta^2 \tQ^4}, \\
        \mathrm{b}_4 &= \frac{4 \left(\left(a^2 \Lambda + 3\right)^2 - 27 \zeta M \tQ^2\right)}{9 \zeta^2 \tQ^4}, \\
        \mathrm{b}_5 &= -\frac{4 \left(a^2 \Lambda - 3\right)}{3 \zeta \tQ^2},
    \end{align}
    \label{eqbj}
\end{subequations}
The two largest positive roots of Eq.~\eqref{eq:rp_eq}, denoted $r_p^-$ and $r_p^+$, correspond to the prograde and retrograde spherical photon orbits, respectively. These characterize the inner and outer boundaries of the photon region surrounding the black hole.

It is essential to highlight that the spherical photon orbits contained within the photon region are crucial in shaping the observed shadow of a black hole. Photons that follow these unstable trajectories, winding around the black hole several times before eventually reaching a distant observer, delineate the edge of the shadow—commonly known as the \emph{critical curve} (see Refs.~\cite{Gralla:2019,bisnovatyi-kogan_analytical_2022,tsupko_shape_2022,claudel_geometry_2001,virbhadra_relativistic_2009,virbhadra_compactness_2024}). A thorough analysis of this boundary for the KNAdS-NLE black hole will be provided in the next section.

\subsection{The black hole shadow}\label{subsec:shadow}

To determine the shadow of the KNAdS–NLE black hole, it is important to note that the spacetime is not asymptotically flat. Consequently, the standard approach of placing the observer at infinity, as in Refs.~\cite{Bardeen:1973a,Bardeen:1973b,Vazquez:2004}, is not applicable. Instead, we position the observer at a finite coordinate location $(r_o,\theta_o)$, equipped with the orthonormal tetrad $\bm{e}_{\{A\}}$ defined as
\begin{eqnarray}
\bm{e}_0 &=& \left.\frac{\left(\Sigma+a^2\sin^2\theta\right)\partial_t+a\,\partial_\phi}{\Xi\sqrt{\Sigma\, \Delta_r}}\right|_{(r_o,\theta_o)}, \label{eq:e0}\\
\bm{e}_1 &=& \left.\sqrt{\frac{\Delta_\theta}{\Sigma}}\, \partial_\theta\right|_{(r_o,\theta_o)}, \label{eq:e1}\\
\bm{e}_2 &=& \left.-\frac{\left(\partial_\phi+a \sin^2\theta\,\partial_t\right)}{\Xi\sqrt{\Sigma } \sin\theta}\right|_{(r_o,\theta_o)}, \label{eq:e2}\\
\bm{e}_3 &=& \left.-\sqrt{\frac{\Delta_r}{\Sigma}}\, \partial_r\right|_{(r_o,\theta_o)}, \label{eq:e3}
\end{eqnarray}
satisfying ${{e}_A}^\alpha {{e}^B}_\alpha = \delta^B_A$.  
Following the method of Ref.~\cite{Grenzebach:2014} (see also \cite{Grenzebach:2016}), this tetrad allows the definition of celestial coordinates in spacetimes with a cosmological background. Here, $\bm{e}_0$ is the observer’s four-velocity, $\bm{e}_3$ points toward the black hole, and $\bm{e}_0 \pm \bm{e}_3$ generate the principal null congruence.

A null geodesic $\bm{\ell}(\tau) = \left(t(\tau),r(\tau),\theta(\tau),\phi(\tau)\right)$, traced backward from the observer toward the black hole, can be expressed in two equivalent ways:
\begin{eqnarray}
\frac{\mathrm{d}\bm{\ell}}{\mathrm{d}\tau} &=& \frac{\mathrm{d} t}{\mathrm{d}\tau} \, \partial_t + \frac{\mathrm{d} r}{\mathrm{d} \tau}\,\partial_r+\frac{\mathrm{d}\theta}{\mathrm{d}\tau}\, \partial_\theta + \frac{\mathrm{d}\phi}{\mathrm{d}\tau}\, \partial_\phi,\\
&=& \mathfrak{c}\left(-\bm{e}_0+\sin\vartheta\cos\psi\,\bm{e}_1 + \sin\vartheta\sin\psi\,\bm{e}_2 + \cos\vartheta\,\bm{e}_3\right),
\end{eqnarray}
where $\vartheta$ and $\psi$ are the celestial coordinates in the observer’s sky, and
\begin{equation}
\mathfrak{c} = \bm{g}\left(\bm{\ell},\bm{e}_0\right) = \frac{a L-\left(\Sigma+a^2\sin^2\theta\right)E}{\Xi\sqrt{\Sigma\,\Delta_r}}.
\end{equation}
The direction $\vartheta = 0$ points exactly toward the black hole. The shadow’s boundary corresponds to null geodesics asymptotically approaching unstable spherical photon orbits at $r_p$, characterized by the critical impact parameters $\xi_p$ and $\eta_p$ given in Eqs.~\eqref{e46} and \eqref{e47}. For an observer at $(r_o,\theta_o)$, the associated celestial coordinates $(\psi_p,\vartheta_p)$ are \cite{Grenzebach:2014}
\begin{eqnarray}
\mathcal{P}(r_p,\theta_o) &\vcentcolon=& \sin\psi_p = \frac{\xi_p+a\cos^2\theta_o-a}{\sqrt{\Delta_\theta\eta_p}\,\sin\theta_o}, \label{eq:psi_p}\\
\mathcal{T}(r_p,r_o) &\vcentcolon=& \sin\vartheta_p = \frac{\sqrt{\Delta_r(r_o)\,\eta_p}}{r_o^2-a (\xi_p-a)}. \label{eq:vartheta_p}
\end{eqnarray}
The maximum and minimum $\vartheta$ occur for $\psi_p = -\pi/2$ and $\psi_p = \pi/2$, respectively.

For $a \neq 0$, we apply a stereographic projection of the celestial sphere $(\psi_p,\vartheta_p)$ onto a plane to obtain the Cartesian coordinates of the shadow as \cite{Grenzebach:2014}
\begin{eqnarray}
X &=& -2 \tan\left(\frac{\vartheta_p}{2}\right)\sin\psi_p, \label{eq:Xp}\\
Y &=& -2 \tan\left(\frac{\vartheta_p}{2}\right)\cos\psi_p. \label{eq:Yp}
\end{eqnarray}
For $\theta_o = \pi/2$ the observer lies in the equatorial plane. In the non-rotating limit ($a = 0$), the photon region reduces to a photon sphere, and the shadow remains perfectly circular due to the preservation of circularity under stereographic projection, regardless of the observer’s motion.

\begin{figure}[t]
    \centering
    \includegraphics[width=7cm]{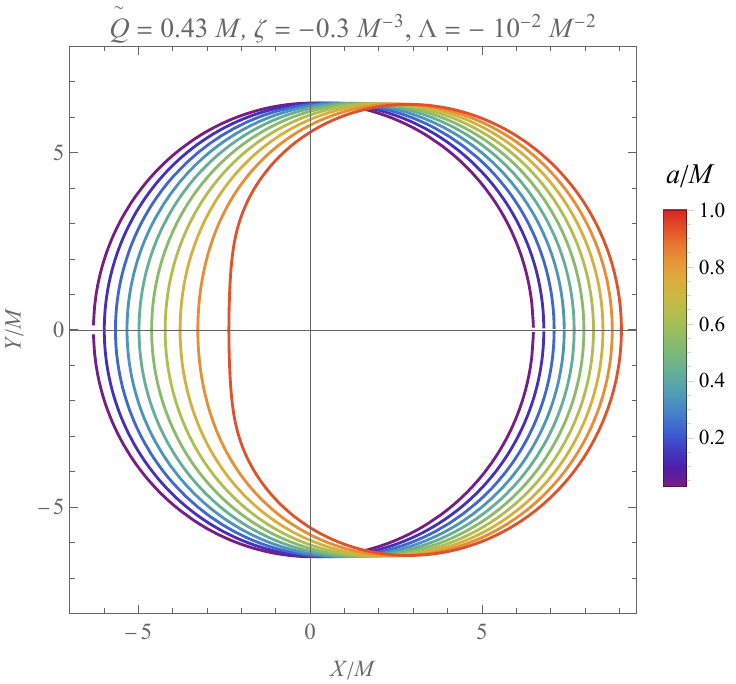} (a)\quad
    \includegraphics[width=7cm]{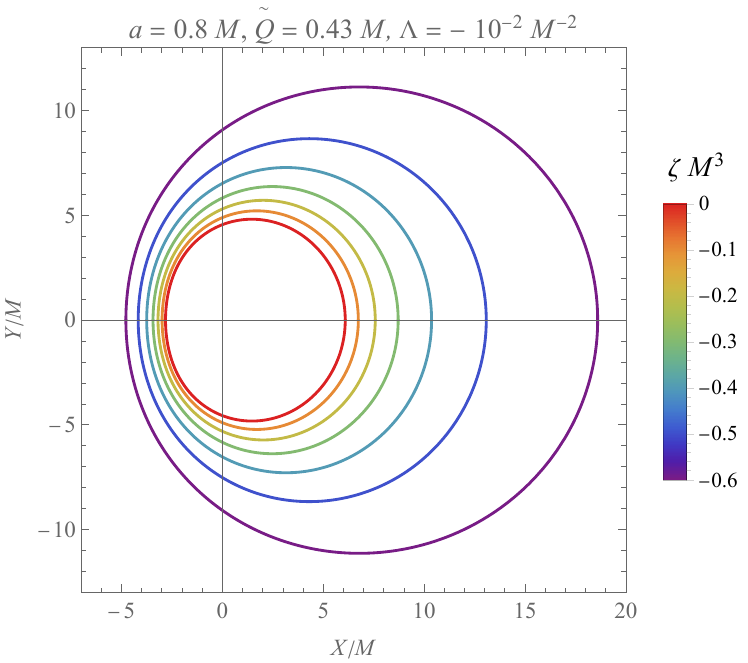} (b)
    \includegraphics[width=7cm]{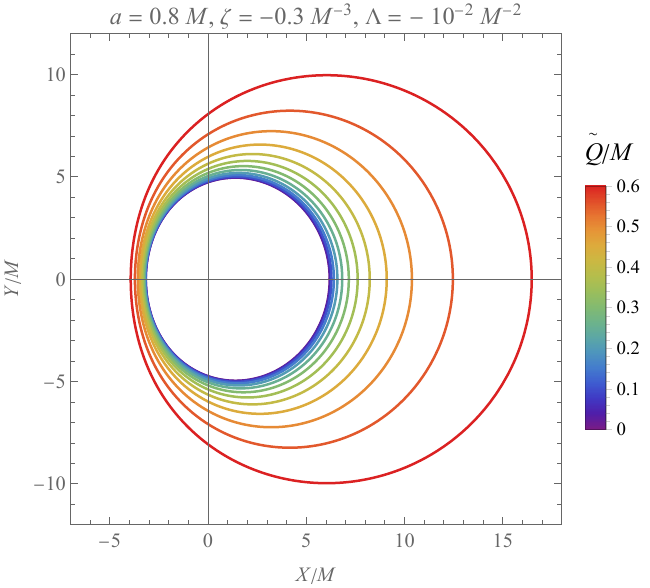} (c)
    \caption{The shadow boundaries of the KNAdS-NLE black hole in the $X$-$Y$ plane for changes in the spacetime parameters, plotted for an observer located at $r_o=100 M$ and $\theta_o=\pi/2$.}
    \label{fig:shadows}
\end{figure}

Using Eqs.~\eqref{eq:Xp} and \eqref{eq:Yp}, Fig.~\ref{fig:shadows} shows the shadows of the KNAdS–NLE black hole for representative values of the spacetime parameters, with $r_p$ serving as the curve parameter.
As observed from the diagrams, as expected, the for fixed $(\tQ,\zeta)$, the shadow boundary becomes more oblate as the spin parameter is increased. This is accompanied by a shift towards the positive segment if the $X$-axis. On the other hand, for fixed $(a,\tQ)$, increase in the negative values of the $\zeta$-parameter, results in less oblateness of the shadow, but at the same time the shadow boundary is stretched towards the positive $X$, and the shadow diameter is increased. Finally, for fixed $(a,\zeta)$, increase in the effective charge $\tQ$, increases the shadow diameter, but the oblateness does not suffer from sever changes. Furthermore, this increase results in a sensible stretching of the shadow towards the positive $X$-axis.

In the next section, we employ the concepts discussed in this section, in order to find confinements for the spacetime parameters based on the recent observations of the EHT.

\section{Shadow observables and black hole parameter estimation}\label{sec:observables}

For a black hole model to be observationally viable, its parameters must be constrained using data from the EHT. The measured images of the supermassive black holes M87* and Sgr~A* are in good agreement with the predictions of the Kerr solution in general relativity. Here, we identify the key observables relevant to the KNAdS–NLE black hole and investigate how they behave within this framework. In particular, we focus on those features whose observational bounds offer meaningful constraints on the fundamental parameters of the spacetime.

\subsection{Distortion}

The distortion parameter quantifies how much the shadow of a rotating black hole deviates from a perfect circle. In other words, it measures the degree of departure from circularity in the shadow’s outline. Following the method introduced in Ref.~\cite{Hioki:2009na}, the linear radius of the shadow is defined as
\begin{eqnarray}\label{Rs}
    R_s = \frac{(X_t - X_r)^2 + Y_t^2}{2\,|X_r - X_t|},
\end{eqnarray}
where $R_s$ represents the radius of a reference circle that tangentially passes through the points $(X_t, Y_t)$, $(X_b, Y_b)$, and $(X_r, 0)$. The coordinates $X$ and $Y$ are those defined in Eqs.~\eqref{eq:Xp} and \eqref{eq:Yp}, while the subscripts $t$, $b$, and $r$ denote the top, bottom, and rightmost points of the shadow, respectively.  

The distortion parameter is then expressed as
\begin{eqnarray}\label{deltas}
    \delta_s = \frac{|X_l - X'_l|}{R_s},
\end{eqnarray}
where $X_l$ and $X'_l$ denote the leftmost boundaries of the shadow and of the reference circle, respectively, both lying along the $-X$ axis. Assuming the shadow is symmetric with respect to the $X$ axis, we set $X_t = X_b = 0$ and $Y_b = -Y_t$. The point $Y_t \equiv Y(r_t)$ is determined from the condition $Y'(r_t)/X'(r_t) = 0$, where $r_t$ is the root of this equation. Similarly, $X_r \equiv X(r_r)$ and $X_l \equiv X(r_l)$ correspond to the two positive real roots of the equation $Y^2(r) = 0$.

\begin{figure}[h]
    \centering
    \includegraphics[width=5.3cm]{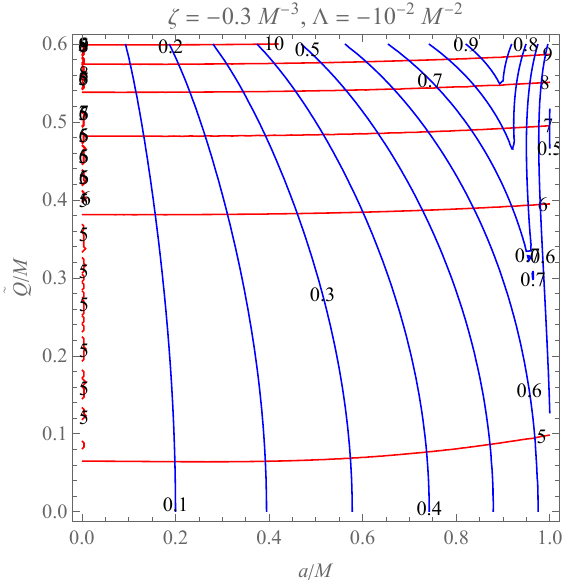} (a)\quad
    \includegraphics[width=5.3cm]{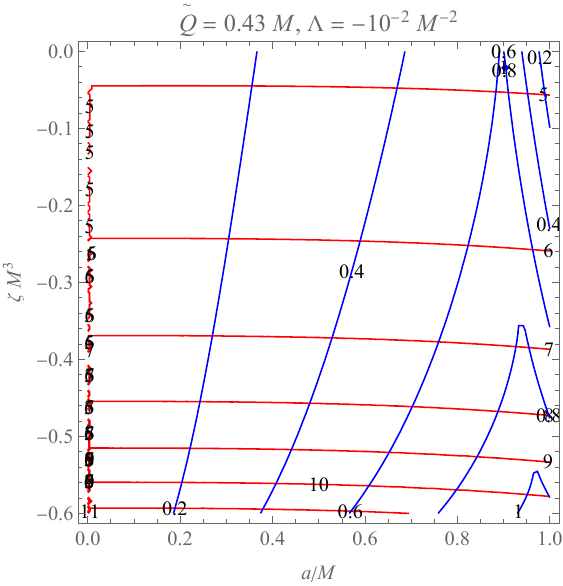} (b)\quad
    \includegraphics[width=5.3cm]{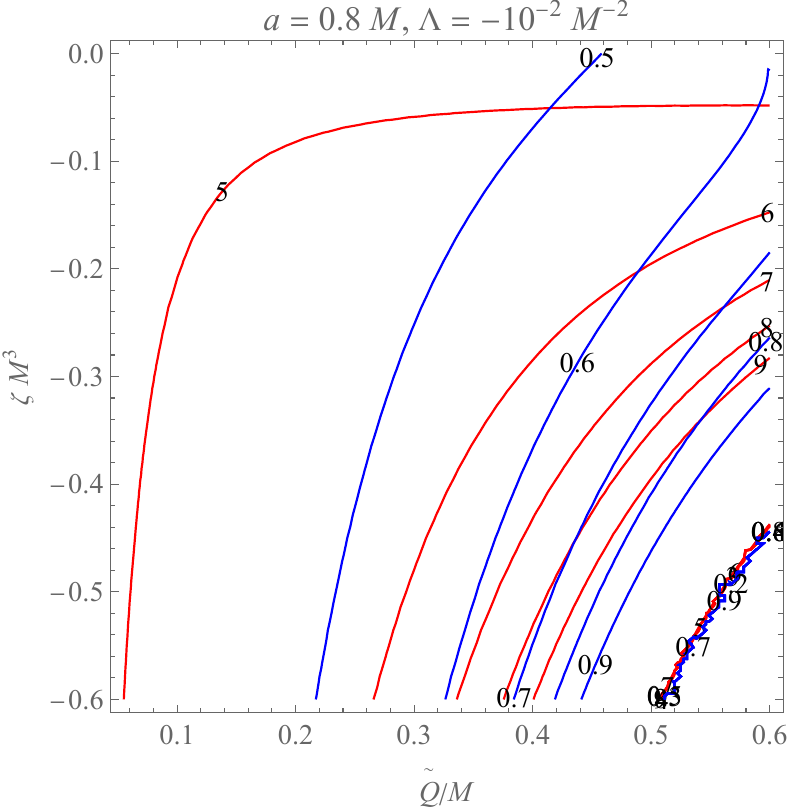} (c)
    \caption{Correlation between $R_s$ (red curves) and $\delta_s$ (blue curves) for an observer at $r_o = 100 M$ and $\theta_o = \pi/4$, shown for various parameter combinations of the KNAdS–NLE black hole.}
    \label{fig:Rs_deltas}
\end{figure}

Table~\ref{tab:3} lists representative values of $R_s$ and $\delta_s$ for the KNAdS–NLE black hole, obtained for different spacetime parameter choices. Figure~\ref{fig:Rs_deltas} illustrates the correlation between $R_s$ and $\delta_s$ in the parameter spaces $(a,\tilde{Q})$, $(a,\zeta)$, and $(\tilde{Q},\zeta)$. This comparison allows for estimating the black hole parameters from the intersection points of $R_s$ and $\delta_s$.
\begin{table}[h]
    \centering
    \begin{tabular}{c|c|c|c|c}
        $R_s/M$ & $\delta_s$ & $a/M$ & $\tilde{Q}/M$ & $\zeta M^3$  \\
         \hline\hline
        5.188 & 0.470 & 0.8 & 0.2 & $-0.3$ \\
        5.700 & 0.560 & 0.8 & 0.43 & $-0.2$ \\
        6.300 & 0.771 & 0.93 & 0.43 & $-0.3$ \\
        6.334 & 0.600 & 0.8 & 0.43 & $-0.3$ \\
        6.372 & 0.427 & 0.6 & 0.43 & $-0.3$ \\
        9.750 & 0.872 & 0.8 & 0.6 & $-0.3$ \\
        10.90 & 0.845 & 0.8 & 0.43 & $-0.6$ \\
    \end{tabular}
    \caption{Representative values of the shadow radius $R_s$ and distortion parameter $\delta_s$ for $\Lambda = -10^{-2} M^{-2}$, $r_o = 100 M$, and $\theta_o = \pi/4$.}
    \label{tab:3}
\end{table}
\subsection{Parameter estimation}

The shadow of a black hole encodes essential information about the geometry of the surrounding spacetime, including its size and shape. This makes it a powerful probe for testing modified gravity theories and for constraining black hole parameters~\citep{jusufi_constraints_2022,okyay_nonlinear_2022,afrin_parameter_2021,chen_superradiant_2022,afrin_tests_2023,sarikulov_shadow_2022,zahid_shadow_2023,zubair_4d_2023}. Parameter estimation can be carried out by analyzing well-defined shadow observables that capture these geometric features.  

Beyond the previously discussed approach, one can employ a coordinate-independent formalism~\citep{abdujabbarov_coordinate-independent_2015,kumar_black_2020} that focuses directly on measurable quantities such as the shadow area and its oblateness. The shadow area, $A_s$, quantifies the overall size of the shadow, while the oblateness, $D_s$, characterizes its departure from circular symmetry. These quantities are defined as~\citep{kumar_black_2020}
\begin{eqnarray}
A_s &=& 2 \int_{r_p^{-}}^{r_p^{+}} Y(r_p)\, X'(r_p) \, \mathrm{d}r_p, \label{Area1} \\
D_s &=& \frac{Y_t - Y_b}{X_r - X_l} = \frac{2 Y_t}{X_r - X_l}, \label{eq:Ds}
\end{eqnarray}
where $(X_t, Y_t)$, $(X_b, Y_b)$, $(X_r, 0)$, and $(X_l, 0)$ denote the characteristic points on the shadow boundary as defined earlier. Note that, for an observer situated in the equatorial plane, the oblateness parameter $D$ can take values in the range $\sqrt{3} \leq D < 1$. The special case $D = 1$ corresponds to the Schwarzschild black hole, while the value $D_s = \sqrt{3}/2$ represents the extremal Kerr black hole limit \cite{Tsupko:2017a}. In this context, $D>1$ corresponds to black hole shadows larger than the Schwarzschild case, characterized by prolateness instead of oblateness. 

\begin{figure}[h]
    \centering
    \includegraphics[width=5.3cm]{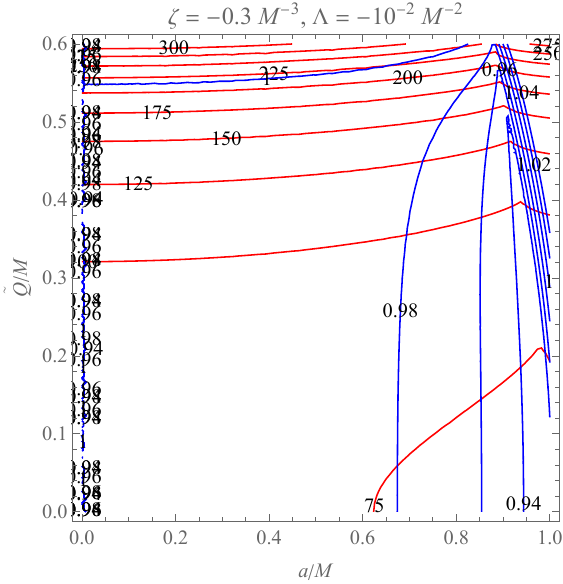} (a)\quad
    \includegraphics[width=5.3cm]{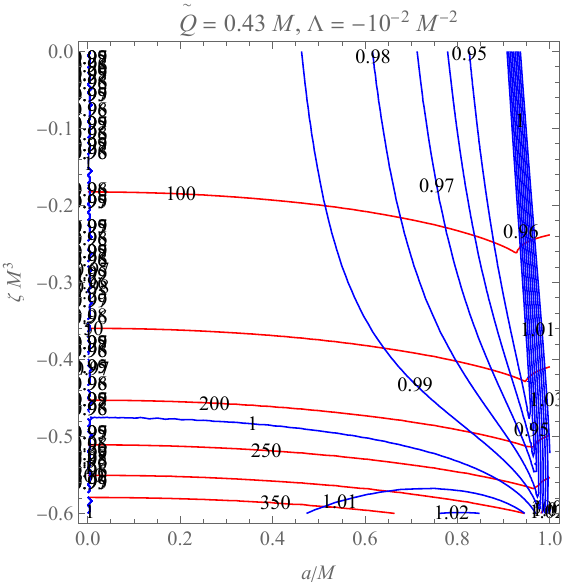} (b)\quad
    \includegraphics[width=5.3cm]{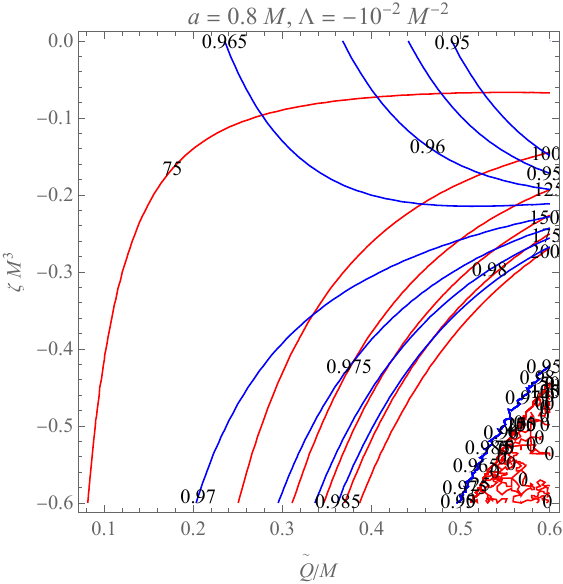} (c)
    \caption{Correlation between $A_s$ (red curves) and $D_s$ (blue curves) for an observer at $r_o = 100 M$ and $\theta_o = \pi/4$, shown for various parameter combinations of the KNAdS–NLE black hole.}
    \label{fig:As_Ds}
\end{figure}

Table~\ref{tab:4} presents representative $A_s$ and $D_s$ values for the KNAdS–NLE black hole under various spacetime parameters. Figure~\ref{fig:As_Ds} shows their correlation in the $(a,\tilde{Q})$, $(a,\zeta)$, and $(\tilde{Q},\zeta)$ spaces, enabling parameter estimation from the intersection points. Furthermore, in Fig~\ref{fig:As_Ds}, we present the relationship between $A_s$ and $D_s$ across the parameter spaces $(a,\tilde{Q})$, $(a,\zeta)$, and $(\tilde{Q},\zeta)$.
\begin{table}[h]
    \centering
    \begin{tabular}{c|c|c|c|c}
        $A_s/M^2$ & $D_s$ & $a/M$ & $\tilde{Q}/M$ & $\zeta M^3$  \\
         \hline\hline
        79.129 & 0.970 & 0.8 & 0.2 & $-0.3$ \\
        93,765 & 0.965 & 0.8 & 0.43 & $-0.2$ \\
        108.445 & 0.930 & 0.93 & 0.43 & $-0.3$ \\
        115.465 & 0.971 & 0.8 & 0.43 & $-0.3$ \\
        121.766 & 0.988 & 0.6 & 0.43 & $-0.3$ \\
        259.646 & 1.004 & 0.8 & 0.6 & $-0.3$ \\
        328.761 & 1.020 & 0.8 & 0.43 & $-0.6$ \\
    \end{tabular}
    \caption{Some exemplary Values of $A_s$ and $D_S$, in accordance with the initial data used in Table \ref{tab:3}.}
    \label{tab:4}
\end{table}
\subsection{Constraints from the EHT observations for M87* and Sgr A*}

The EHT collaboration has recently imaged the shadows of the supermassive black holes M87* and Sgr A* \citep{EventHorizonTelescope:2019dse,EventHorizonTelescope:2019ggy,EventHorizonTelescope:2022wkp,EventHorizonTelescope:2022xqj}, providing an unprecedented opportunity to test theoretical black hole models. These shadow measurements serve as a direct probe of the spacetime geometry and allow one to place observational constraints on black hole parameters within various extensions of general relativity.

In this work, we employ the EHT results for M87* and Sgr A* to constrain the parameters of our rotating black hole solution. The relevant observable is the angular diameter of the shadow, which for a source located at distance $d$ is given by \citep{afrin_parameter_2021,zahid_shadow_2023} 
\begin{equation}
    \theta_d = 2 \frac{R_a}{d}, 
    \qquad R_a = \sqrt{\frac{A_s}{\pi}},
    \label{eq:thetad}
\end{equation}
where $R_a$ denotes the areal radius of the shadow, defined in terms of the shadow area $A_s$. The angular diameter thus depends not only on the black hole parameters but also on the observer’s inclination angle, with an implicit dependence on the black hole mass.  
For M87*, we take the mass and distance to be $M = 6.5 \times 10^9 M_\odot$ and $d = 16.8 \,\text{Mpc}$, respectively \cite{EventHorizonTelescope:2019pgp,EventHorizonTelescope:2019ggy}. The EHT reports an angular diameter of $\theta_d = (42 \pm 3)\,\mu\text{as}$ at the $1\sigma$ confidence level \cite{EventHorizonTelescope:2019dse}. In our analysis, uncertainties in $M$ and $d$ are neglected for simplicity. Furthermore, according to the EHT constraints, the spin of M87* is approximately $a \approx (0.9 \pm 0.05)M$ \cite{Tamburini:2019vrf}, with an inclination $\theta_o \simeq 17^\circ$ \cite{Daly:2023axh}.

A similar analysis can be performed for the supermassive black hole Sgr A*. According to the EHT collaboration, its shadow diameter is measured as $\theta_d = (48.7 \pm 7)\,\mu\text{as}$ \cite{EventHorizonTelescope:2022wkp}. The mass and distance are taken to be $M = 4 \times 10^6 M_\odot$ and $d = 8\,\text{kpc}$, respectively \cite{EventHorizonTelescope:2022exc,EventHorizonTelescope:2022xqj}. Moreover, recent analyses suggest that the spin of Sgr A* is approximately $a = (0.9 \pm 0.06)M$ \citep{Daly:2023axh}.

\begin{figure}[h]
    \centering
    \includegraphics[width=8cm]{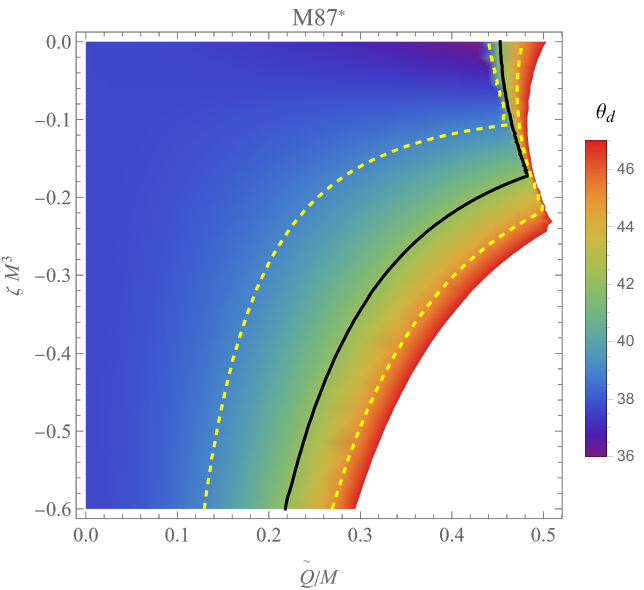} (a)\qquad
    \includegraphics[width=8cm]{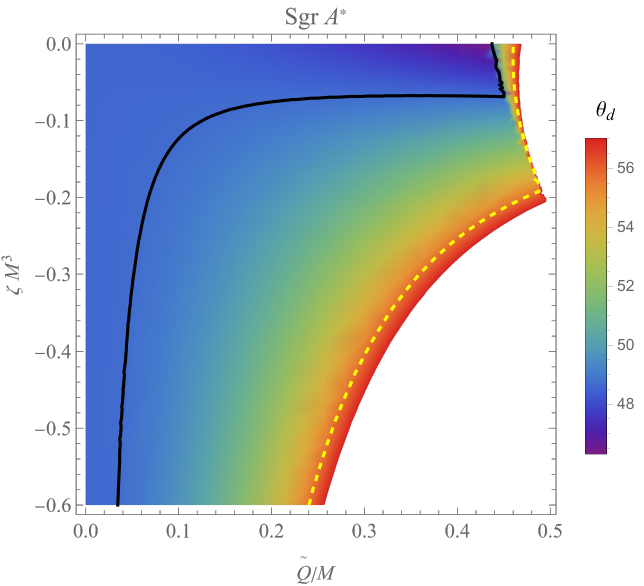} (b)
    \caption{Angular diameter $\theta_d$ of the KNAdS-NLE black hole shadows as a function of $\tilde{Q}$ and $\zeta$ at inclination $17^\circ$, $r_o=d$ for each of the black holes with $a = 0.9M$, and $\Lambda = -10^{-52} M^2$. Panel (a) corresponds to M87*, where the solid black curve denotes $\theta_d = 42\,\mu\text{as}$ and the dashed yellow curves represent the $\pm1\sigma$ uncertainty of $\pm 3\,\mu\text{as}$ reported by the EHT. Panel (b) shows the case of Sgr A*, with the solid black curve indicating $\theta_d = 48.7\,\mu\text{as}$ and the dashed yellow curve marking the $+1\sigma$ uncertainty of $+ 7\,\mu\text{as}$.}
    \label{fig:EHTQzeta}
\end{figure}

Figure~\ref{fig:EHTQzeta} displays the theoretical prediction for $\theta_d$ as a function of the effective charge $\tQ$ and the nonlinear coefficient $\zeta$, for $a=0.9 M$ and $\theta_o = 17^\circ$, within the EHT observations.

From the results presented in Fig.~\ref{fig:EHTQzeta}(a), the constraints from the EHT observations of M87* indicate that, for the KNAdS black hole with $\zeta = 0$, the effective charge lies in the interval $0.44 M \leq \tQ \leq 0.48 M$. As the nonlinearity parameter $\zeta$ decreases, this interval contracts, reaching its narrowest range at $\zeta = -0.1 M^{-3}$, where the allowed values are $0.46 M \leq \tQ \leq 0.47 M$. Beyond this point, further decrease in $\zeta$ enlarges the allowed interval, with its widest range attained at $\zeta = -0.21 M^{-3}$, corresponding to $0.24 M \leq \tQ \leq 0.50 M$.

For the case of Sgr A*, shown in Fig.~\ref{fig:EHTQzeta}(b), the KNAdS black hole yields the range $0.44 M \leq \tQ \leq 0.460 M$ when $\zeta = 0$. The minimum interval occurs at $\zeta = -0.06 M^{-3}$, giving $0.45 M \leq \tQ \leq 0.462 M$. Conversely, the maximum allowed interval appears at $\zeta = -0.19 M^{-3}$, where the effective charge spans $0.07 M \leq \tQ \leq 0.49 M$.

Taken together, the EHT constraints for both M87* and Sgr A* suggest that the most consistent and observationally reliable range for the effective charge lies approximately within
\begin{equation}
    0.44 M \lesssim \tQ \lesssim 0.48 M,
    \label{eq:tQ_range}
\end{equation}
with small variations depending on the value of the nonlinearity parameter $\zeta$.


\section{Energy emission rate}\label{sec:EnergyEmission}

\begin{figure}[h]
    \centering
    \includegraphics[width=5.3cm]{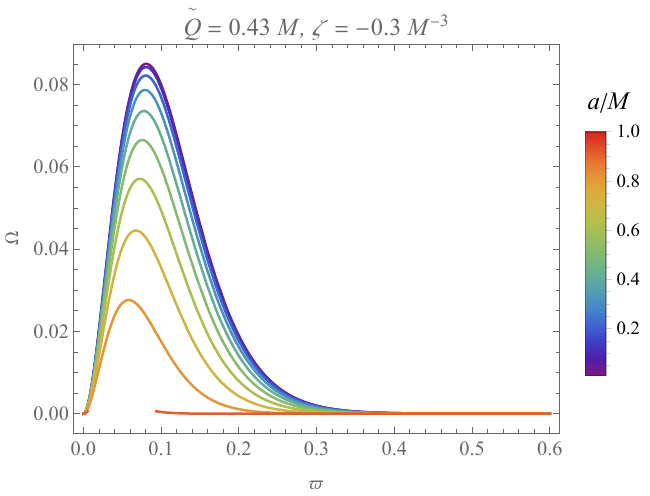} (a)\quad
    \includegraphics[width=5.3cm]{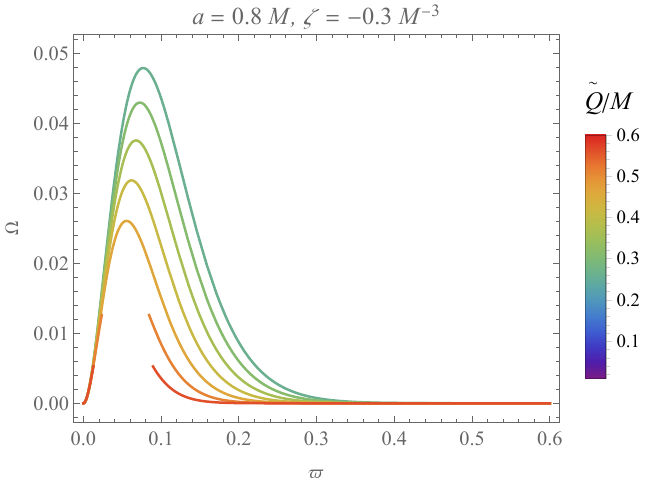} (b)\quad
    \includegraphics[width=5.3cm]{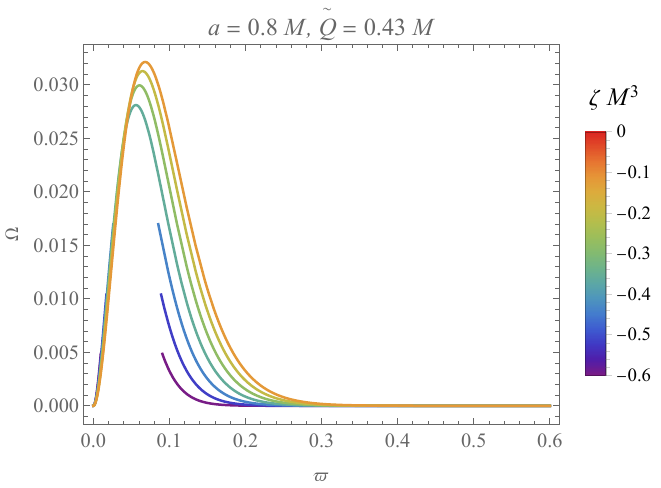} (c)
    \caption{Variation of the energy emission rate with frequency $\varpi$ for different sets of black hole parameters, by assuming $\theta_o=17^\circ$, $r_o=100 M$ and $\Lambda=-10^{-2} M^{-2}$.}
    \label{fig:Omega}
\end{figure}

From a quantum mechanical perspective, particle creation and annihilation can occur near the event horizon of a black hole. In this process, particles with positive energy may tunnel through the horizon and escape, leading to the emission of radiation that gradually reduces the black hole’s mass and energy, eventually resulting in its evaporation. This effect, known as Hawking radiation, originates from quantum fluctuations and causes black holes to radiate thermally \cite{Hawking:1974rv}. At high energies, Hawking radiation is associated with a finite cross-sectional area, denoted by $\sigma_l$, which for distant observers asymptotically approaches the shadow of the black hole \cite{belhaj_deflection_2020, wei_observing_2013}. The quantity $\sigma_l$ is closely linked to the photon ring area and can be approximated as \cite{wei_observing_2013, decanini_fine_2011, li_shadow_2020}
\begin{equation}
    \sigma_l \approx \pi R_{s}^2.
    \label{eq:sigmal}
\end{equation}
Accordingly, the energy emission rate of the black hole takes the form
\begin{equation}
    \Omega \equiv \frac{\mathrm{d}^2 E(\varpi)}{\mathrm{d}\varpi \, \mathrm{d}t} = \frac{2\pi^2 \sigma_l}{e^{\varpi / T_\mathrm{H}^+} - 1} \, \varpi^3 \approx \frac{2\pi^3 R_{s}^2 \varpi^3}{e^{\varpi / T_\mathrm{H}^+} - 1},
    \label{eq:emissionrate}
\end{equation}
where $\varpi$ is the emission frequency and $T_\mathrm{H}^+ = \tilde{\kappa}_g / (2\pi)$ denotes the Hawking temperature at the event horizon. The surface gravity $\tilde{\kappa}_g$ is expressed as
\begin{equation}
    \tilde{\kappa}_g = \left. \frac{\Delta_r'(r)}{2 \left(a^2 + r^2\right)} \right|_{r_+},
    \label{eq:kappa}
\end{equation}
with $r_+$ being the horizon radius. In the special case of vanishing spin ($a = 0$), this reduces to $\tilde{\kappa}_g = f'(r_+)/2$, which corresponds to the standard surface gravity of static black holes. Figure~\ref{fig:Omega} illustrates the behavior of $\Omega$ as a function of frequency $\varpi$ for the KNAdS-NLE black hole model.
From the diagrams, it is evident that increasing the spin parameter, the effective charge, and the nonlinearity parameter consistently leads to a reduction in the emission rate.

\section{Conclusions}\label{sec:conclusion}

In this paper, we have studied the optical and thermodynamic properties of the Kerr-Newman-Anti-de Sitter black hole coupled to a nonlinear electrodynamics (KNAdS-NLE black hole). By employing the tetrad formalism of celestial coordinates, we constructed the shadow for observers at finite distance, which is necessary due to the non-asymptotically flat nature of the geometry. This approach enabled us to compute and analyze key shadow observables, including the radius, distortion parameter, area, and oblateness, and to evaluate their dependence on the spin parameter, the effective charge, and the nonlinearity parameter.

Our results show that the interplay between charge and nonlinearity introduces distinctive signatures in the shadow profile. While the spin drives the departure from circularity through increased distortion and oblateness, the nonlinear electrodynamics modifies both the size and shape of the shadow in a nontrivial manner. These effects are particularly relevant when comparing theoretical predictions with high-resolution black hole imaging.

We further constrained the parameter space of the KNAdS-NLE solution using the EHT observations of the supermassive black holes M87* and Sgr A*. By relating the areal radius of the shadow to its angular diameter, we derived bounds on the effective charge and nonlinearity parameter across different symmetry groups. For M87*, the observational consistency favors narrow ranges of the effective charge that depend on the degree of nonlinearity, while for Sgr A*, broader intervals are allowed. Importantly, our analysis shows that increasing the nonlinearity can either shrink or expand the viable parameter space, leading to a rich phenomenology that can be tested with future observations.

In addition, we investigated the Hawking radiation of the KNAdS-NLE black hole through its energy emission rate. The analysis indicates that higher spin, larger effective charge, and stronger nonlinearity all contribute to suppressing the emission spectrum. This feature is consistent with the expectation that nonlinear electromagnetic fields may act as stabilizing agents in near-horizon physics, thereby affecting the evaporation process of black holes.

Taken together, these findings suggest that the KNAdS-NLE solution provides a compelling framework to test extensions of general relativity and to interpret black hole images within a more general theoretical context. Future improvements in very-long-baseline interferometry (VLBI) and multi-wavelength EHT observations will allow for sharper constraints on deviations from the Kerr paradigm, potentially distinguishing between standard Einstein-Maxwell black holes and models involving nonlinear electrodynamics. Moreover, the generalization to rotating solutions via the Newman-Janis algorithm, combined with strong lensing analyses, remains a promising avenue for further research.

In summary, the KNAdS-NLE black hole exhibits a set of observational and theoretical features that make it both astrophysically relevant and mathematically rich. Its shadow characteristics, consistency with EHT data, and quantum emission properties highlight its potential as a laboratory for probing gravitational physics beyond general relativity.

\begin{acknowledgments}
This work has been supported by Universidad Central de Chile through project No. PDUCEN20240008. 
\end{acknowledgments}

\section*{Data Availability}

No original or third-party data were generated or analyzed in this work.

\appendix

\section{An analytical method to obtain the roots of quartics}\label{app:A}

The quartic equation \eqref{eq:Deltar=0} can be depressed by using the change of variable $r\doteq x-\mathrm{a}_3/4$, which changes it to the form
\begin{equation}
 x^4 + \mathcal{a} x^2 + \mathcal{b} x + \mathcal{c} = 0,
    \label{eq:A1}
\end{equation}
where
\begin{subequations}
    \begin{align}
        & \mathcal{a} = \mathrm{a}_2-\frac{3 \mathrm{a}_3^2}{8},\\
        & \mathcal{b} = \mathrm{a}_1 + \frac{\mathrm{a}_3^3}{8} -\frac{\mathrm{a}_3\mathrm{a}_2}{2},\\
        & \mathcal{c} = \mathrm{a}_0 + \frac{\mathrm{a}_3^2 \mathrm{a}_2}{16} -\frac{3\mathrm{a}_3^4}{256} - \frac{\mathrm{a}_3\mathrm{a}_1}{4}.
    \end{align}
    \label{eq:A2}
\end{subequations}
This equation can be recast as the product of two quadratic equations, as below:
\begin{equation}
x^4 + \mathcal{a} x^3 + \mathcal{b} x^2 + \mathcal{c} = \bigl(x^2-2\tilde{\alpha}x+\tilde{\beta}\bigr) \bigl(x^2+2\tilde{\alpha}x+\tilde{\gamma}\bigr),
    \label{eq:A3}
\end{equation}
from which we obtain
\begin{subequations}
\begin{align}
&\mathcal{a}=\tilde\beta+\tilde\gamma-4\tilde\alpha^2,  \\
&\mathcal{b}=2\tilde\alpha(\tilde\beta-\tilde\gamma),\\
&\mathcal{c}=\tilde\beta \tilde\gamma\label{eq:app_Abeta}.
\end{align}
\label{eq:A4}
\end{subequations}
Now by solving the first two equations for $\tilde\beta$ and $\tilde\gamma$, we get to the expressions
\begin{subequations}
\begin{align}
&\tilde\beta=2\tilde\alpha^2 +{\mathcal{a}\over 2}+{\mathcal{b}\over 4\tilde\alpha},\\
&\tilde\gamma=2\tilde\alpha^2 +{\mathcal{a}\over 2}-{\mathcal{b}\over 4\tilde\alpha},
\end{align}
\label{eq:A5}
\end{subequations}
which together with Eq.~\eqref{eq:app_Abeta}, provides the following sextic polynomial equation for $\tilde\alpha$:
\begin{equation}
\tilde\alpha^6+{\mathcal{a}\over 2}\tilde\alpha^4+\left( {\mathcal{a}^2\over 16}-{\mathcal{c}\over 4}\right) \tilde\alpha^2-{\mathcal{b}^2\over 64}=0.
\label{eq:A6}
\end{equation}
Now applying the change of variable
\begin{equation}
    \tilde\alpha^2=\tilde{U}-{\mathcal{a}\over 6},
    \label{eq:A7}
\end{equation}
we obtain the depressed cubic equation
\begin{equation}
\tilde{U}^3-\tilde\eta_2 \,\tilde{U}-\tilde\eta_3=0,
\label{eq:A8}
\end{equation}
where
\begin{subequations}
\begin{align}
&\tilde\eta_2={\mathcal{a}^2\over 48}+{\mathcal{c}\over 4},\\
&\tilde\eta_3={\mathcal{a}^3\over 864}+{\mathcal{b}^2\over 64}-{\mathcal{a} \mathcal{c}\over 24}.
\end{align}
\label{eq:A9}
\end{subequations}
The real solution to this cubic equation is obtained as \cite{Nickalls:2006,Zucker:2008} 
\begin{equation}
\tilde{U}= 2\sqrt{{\tilde\eta_2\over 3}} \cosh\left(\frac{1}{3}\arccosh\left({3\over 2}\tilde\eta_3\sqrt{{3\over \tilde\eta_2^3}}  \right) \right).
\label{eq:A10}
\end{equation}
Therefore, the roots of Eq.~\eqref{eq:A1} are 
\begin{eqnarray}
&& x_1=\tilde\alpha+\sqrt{\tilde\alpha^2-\tilde\beta},\label{eq:A11}\\
&& x_2=\tilde\alpha-\sqrt{\tilde\alpha^2-\tilde\beta},\label{eq:A12}\\
&&  x_3=-\tilde\alpha+\sqrt{\tilde\alpha^2-\tilde\gamma},\label{eq:A13}\\
&&  x_4=-\tilde\alpha-\sqrt{\tilde\alpha^2-\tilde\gamma},\label{eq:A14}
\end{eqnarray}
which provide the roots of the quartic \eqref{eq:Deltar=0} as
\begin{equation}
     r_j = x_j - \frac{\mathrm{a}_3}{4},\qquad j = \overline{1,4}. 
\end{equation}

\section{The Kretschmann scalar for the KNAdS-NLE black hole spacetime}\label{app:B}

The Kretschmann scalar is calculated as
\begin{multline}
\mathcal{K} = R_{\mu\nu\rho\sigma}R^{\mu\nu\rho\sigma} = \frac{1}{12 \left[a^2 \cos (2 \theta )+a^2+2 r^2\right]^6}\Biggl\{
-11520 a^6 M^2 + 16128 a^4 \tQ^4 - 138240 a^4 M \tQ^2 r + 
 207360 a^4 M^2 r^2 \\
 - 104448 a^2 \tQ^4 r^2 + 368640 a^2 M \tQ^2 r^3 - 
 276480 a^2 M^2 r^4 + 43008 \tQ^4 r^4 - 73728 M \tQ^2 r^5
 + 36864 M^2 r^6\\
 -5760 a^6 \zeta  \tQ^4 r+69120 a^6 \zeta  M \tQ^2 r^2+149760 a^4 \zeta  \tQ^4 r^3-276480 a^4 \zeta  M \tQ^2 r^4-89088 a^2 \zeta  \tQ^4 r^5+110592 a^2 \zeta  M \tQ^2 r^6\\
 -6144 \zeta  \tQ^4 r^7+7560 a^8 \zeta ^2 \tQ^4 r^2-31680 a^6 \zeta ^2 \tQ^4 r^4+61056 a^4 \zeta ^2 \tQ^4 r^6+1536 a^2 \zeta ^2 \tQ^4 r^8+6144 \zeta ^2 \tQ^4 r^{10}\\
 -1512 a^{10} \zeta  \Lambda  \tQ^2 r-8400 a^8 \zeta  \Lambda  \tQ^2 r^3-19200 a^6 \zeta  \Lambda  \tQ^2 r^5-23040 a^4 \zeta  \Lambda  \tQ^2 r^7-15360 a^2 \zeta  \Lambda  \tQ^2 r^9-6144 \zeta  \Lambda  \tQ^2 r^{11}\\
 +462 a^{12} \Lambda ^2+3024 a^{10} \Lambda ^2 r^2+8400 a^8 \Lambda ^2 r^4+12800 a^6 \Lambda ^2 r^6+11520 a^4 \Lambda ^2 r^8+6144 a^2 \Lambda ^2 r^{10}+2048 \Lambda ^2 r^{12}\\
 + 24 a^2 \cos (2 \theta ) \Biggl[33 a^{10} \Lambda ^2+105 a^8 \Lambda  r \left(2 \Lambda  r-\zeta  \tQ^2\right)+56 a^6 r^2 \left(9 \zeta ^2 \tQ^4-10 \zeta  \Lambda  \tQ^2 r+10 \Lambda ^2 r^2\right)-20 a^4 \Big(36 M^2\\
-216 \zeta  M \tQ^2 r^2+9 \zeta  \tQ^4 r \left(11 \zeta  r^3+2\right)+60 \zeta  \Lambda  \tQ^2 r^5-40 \Lambda ^2 r^6\Big)+64 a^2 \Big(10 r^2 \left(18 M^2+\Lambda ^2 r^6\right)-20 \tQ^2 \bigl[6 M \left(2 \zeta  r^4+r\right)\\
+\zeta  \Lambda  r^7\bigr]+\tQ^4 \left(53 \zeta ^2 r^6+130 \zeta  r^3+14\right)\Big)+64 r^2 \Big(-180 M^2 r^2+2 \tQ^2 \bigl[12 M r \left(3 \zeta  r^3+10\right)-5 \zeta  \Lambda  r^7\bigr]+\tQ^4 \left(\zeta ^2 r^6-58 \zeta  r^3-68\right)\\
+4 \Lambda ^2 r^8\Big)\Biggr]
+3 a^4 \cos (4 \theta ) \Biggl[165 a^8 \Lambda ^2+480 a^6 \Lambda  r \left(2 \Lambda  r-\zeta  \tQ^2\right)+224 a^4 r^2 \left(9 \zeta ^2 \tQ^4-10 \zeta  \Lambda  \tQ^2 r+10 \Lambda ^2 r^2\right)\\
-64 a^2 \Big(36 M^2-216 \zeta  M \tQ^2 r^2
+9 \zeta  \tQ^4 r \left(11 \zeta  r^3+2\right)+60 \zeta  \Lambda  \tQ^2 r^5-40 \Lambda ^2 r^6\Big)+128 \Big(10 r^2 \left(18 M^2+\Lambda ^2 r^6\right)\\-20 \tQ^2 \left(6 M \left(2 \zeta  r^4+r\right)+\zeta  \Lambda  r^7\right)
+\tQ^4 \left(53 \zeta ^2 r^6+130 \zeta  r^3+14\right)\Big)\Biggr]
+220 a^{12} \Lambda ^2 \cos (6 \theta )+66 a^{12} \Lambda ^2 \cos (8 \theta )\\
+12 a^{12} \Lambda ^2 \cos (10 \theta )+a^{12} \Lambda ^2 \cos (12 \theta )-540 a^{10} \zeta  \Lambda  \tQ^2 r \cos (6 \theta )-120 a^{10} \zeta  \Lambda  \tQ^2 r \cos (8 \theta )-12 a^{10} \zeta  \Lambda  \tQ^2 r \cos (10 \theta )\\
+1080 a^{10} \Lambda ^2 r^2 \cos (6 \theta )+240 a^{10} \Lambda ^2 r^2 \cos (8 \theta )+24 a^{10} \Lambda ^2 r^2 \cos (10 \theta )+1728 a^8 \zeta ^2 \tQ^4 r^2 \cos (6 \theta )
+216 a^8 \zeta ^2 \tQ^4 r^2 \cos (8 \theta )\\
-1920 a^8 \zeta  \Lambda  \tQ^2 r^3 \cos (6 \theta )-240 a^8 \zeta  \Lambda  \tQ^2 r^3 \cos (8 \theta )+1920 a^8 \Lambda ^2 r^4 \cos (6 \theta )+240 a^8 \Lambda ^2 r^4 \cos (8 \theta )-1152 a^6 M^2 \cos (6 \theta )\\
+6912 a^6 \zeta  M \tQ^2 r^2 \cos (6 \theta )-3168 a^6 \zeta ^2 \tQ^4 r^4 \cos (6 \theta )-576 a^6 \zeta  \tQ^4 r \cos (6 \theta )-1920 a^6 \zeta  \Lambda  \tQ^2 r^5 \cos (6 \theta )\\
+1280 a^6 \Lambda ^2 r^6 \cos (6 \theta )
\Biggr\}.
    \label{eq:B1}
\end{multline}

\bibliographystyle{ieeetr}
\bibliography{biblio_v1}

\begin{thebibliography}{10}

\bibitem{LIGOScientific:2016aoc}
B.~P. Abbott {\em et~al.}, ``{Observation of Gravitational Waves from a Binary Black Hole Merger},'' {\em Phys. Rev. Lett.}, vol.~116, no.~6, p.~061102, 2016.

\bibitem{abbott_gwtc-3:_2023}
R.~Abbott {\em et~al.}, ``{GWTC}-3: {Compact} {Binary} {Coalescences} {Observed} by {LIGO} and {Virgo} during the {Second} {Part} of the {Third} {Observing} {Run},'' {\em Physical Review X}, vol.~13, p.~041039, Dec. 2023.

\bibitem{ayon-beato_regular_1998}
E.~Ayón-Beato and A.~García, ``Regular {Black} {Hole} in {General} {Relativity} {Coupled} to {Nonlinear} {Electrodynamics},'' {\em Physical Review Letters}, vol.~80, pp.~5056--5059, June 1998.

\bibitem{breton_rotating_2022}
N.~Bretón, C.~Lämmerzahl, and A.~Macías, ``Rotating structure of the {Euler}-{Heisenberg} black hole,'' {\em Physical Review D}, vol.~105, p.~104046, May 2022.

\bibitem{plebanski_electromagnetic_1960}
J.~Plebanski, ``Electromagnetic {Waves} in {Gravitational} {Fields},'' {\em Physical Review}, vol.~118, pp.~1396--1408, June 1960.

\bibitem{boillat_nonlinear_1970}
G.~Boillat, ``Nonlinear {Electrodynamics}: {Lagrangians} and {Equations} of {Motion},'' {\em Journal of Mathematical Physics}, vol.~11, pp.~941--951, Mar. 1970.

\bibitem{novello_geometrical_2000}
M.~Novello, V.~A. De~Lorenci, J.~M. Salim, and R.~Klippert, ``Geometrical aspects of light propagation in nonlinear electrodynamics,'' {\em Physical Review D}, vol.~61, p.~045001, Jan. 2000.

\bibitem{obukhov_fresnel_2002}
Y.~N. Obukhov and G.~F. Rubilar, ``Fresnel analysis of wave propagation in nonlinear electrodynamics,'' {\em Physical Review D}, vol.~66, p.~024042, July 2002.

\bibitem{schellstede_causality_2016}
G.~O. Schellstede, V.~Perlick, and C.~Lämmerzahl, ``On causality in nonlinear vacuum electrodynamics of the {Plebański} class,'' {\em Annalen der Physik}, vol.~528, pp.~738--749, Oct. 2016.

\bibitem{witten_anti_1998}
E.~Witten, ``Anti {De} {Sitter} {Space} {And} {Holography},'' 1998.

\bibitem{Garcia-Diaz:2021bao}
A.~A. {Garcia-Diaz}, ``{Stationary Rotating Black Hole Exact Solution within Einstein--Nonlinear Electrodynamics},'' {\em arXiv e-prints}, p.~arXiv:2112.06302, Dec. 2021.

\bibitem{garcia-diaz_adsds_2022}
A.~A. García-Díaz, ``{AdS}–{dS} stationary rotating black hole exact solution within {Einstein}-nonlinear electrodynamics,'' {\em Annals of Physics}, vol.~441, p.~168880, June 2022.

\bibitem{galindo-uriarte_nonlinear_2024}
O.~Galindo-Uriarte and N.~Breton, ``Nonlinear electromagnetic generalization of the {Kerr}-{Newman} solution with a cosmological constant,'' {\em Physical Review D}, vol.~110, p.~064021, Sept. 2024.

\bibitem{EventHorizonTelescope:2019pgp}
K.~Akiyama {\em et~al.}, ``{First M87 Event Horizon Telescope Results. V. Physical Origin of the Asymmetric Ring},'' {\em Astrophys. J. Lett.}, vol.~875, no.~1, p.~L5, 2019.

\bibitem{EventHorizonTelescope:2022wkp}
K.~Akiyama {\em et~al.}, ``{First Sagittarius A* Event Horizon Telescope Results. I. The Shadow of the Supermassive Black Hole in the Center of the Milky Way},'' {\em Astrophys. J. Lett.}, vol.~930, no.~2, p.~L12, 2022.

\bibitem{filho_analysis_2025}
A.~Araújo~Filho, ``Analysis of a nonlinear electromagnetic generalization of the {Reissner}–{Nordström} black hole,'' {\em The European Physical Journal C}, vol.~85, p.~454, Apr. 2025.

\bibitem{araujo_filho_remarks_2025}
A.~Araújo~Filho, ``Remarks on a nonlinear electromagnetic extension in {AdS} {Reissner}-{Nordström} spacetime,'' {\em Journal of Cosmology and Astroparticle Physics}, vol.~2025, p.~072, Jan. 2025.

\bibitem{penrose_extraction_1971}
R.~Penrose and R.~M. Floyd, ``Extraction of {Rotational} {Energy} from a {Black} {Hole},'' {\em Nature Physical Science}, vol.~229, pp.~177--179, Feb. 1971.

\bibitem{pugliese_observers_2018}
D.~Pugliese and H.~Quevedo, ``Observers in {Kerr} spacetimes: the ergoregion on the equatorial plane,'' {\em The European Physical Journal C}, vol.~78, p.~69, Jan. 2018.

\bibitem{Carter:1968}
B.~Carter, ``Global structure of the kerr family of gravitational fields,'' {\em Physical Review}, vol.~174, pp.~1559--1571, Oct 1968.

\bibitem{Chandrasekhar:2002}
S.~Chandrasekhar, {\em The mathematical theory of black holes}.
\newblock Oxford classic texts in the physical sciences, Oxford University Press, 2002.

\bibitem{Slany:2020jhs}
P.~Slan\'y and Z.~Stuchl\'\i{}k, ``{Equatorial circular orbits in Kerr\textendash{}Newman\textendash{}de Sitter spacetimes},'' {\em Eur. Phys. J. C}, vol.~80, no.~6, p.~587, 2020.

\bibitem{PhysRevD.81.044020}
E.~Hackmann, C.~L\"ammerzahl, V.~Kagramanova, and J.~Kunz, ``Analytical solution of the geodesic equation in kerr-(anti-) de sitter space-times,'' {\em Phys. Rev. D}, vol.~81, p.~044020, 2010.

\bibitem{StuchlikEPJC2018}
Z.~Stuchl\'{i}k, D.~Charbul\'{a}k, and J.~Schee, ``Light escape cones in local reference frames of kerr–de sitter black hole spacetimes and related black hole shadows,'' {\em Eur. Phys. J. C}, vol.~78, 2018.

\bibitem{Carter:1968rr}
B.~Carter, ``{Global structure of the Kerr family of gravitational fields},'' {\em Phys. Rev.}, vol.~174, pp.~1559--1571, 1968.

\bibitem{Bardeen:1972a}
J.~M. {Bardeen}, W.~H. {Press}, and S.~A. {Teukolsky}, ``{Rotating Black Holes: Locally Nonrotating Frames, Energy Extraction, and Scalar Synchrotron Radiation},'' {\em Astrophysical Journal}, vol.~178, pp.~347--370, Dec. 1972.

\bibitem{Bardeen:1973b}
J.~Bardeen, ``{Timelike and null geodesics in the Kerr metric},'' in {\em {Les Houches Summer School of Theoretical Physics}: {Black Holes}}, pp.~215--240, 1973.

\bibitem{stoghianidis_polar_1987}
E.~Stoghianidis and D.~Tsoubelis, ``Polar orbits in the {Kerr} space-time,'' {\em General Relativity and Gravitation}, vol.~19, pp.~1235--1249, Dec. 1987.

\bibitem{cramer_using_1997}
C.~R. Cramer, ``Using the {Uncharged} {Kerr} {Black} {Hole} as a {Gravitational} {Mirror},'' {\em General Relativity and Gravitation}, vol.~29, pp.~445--454, Apr. 1997.

\bibitem{Teo:2003}
E.~Teo, ``Spherical {Photon} {Orbits} {Around} a {Kerr} {Black} {Hole},'' {\em General Relativity and Gravitation}, vol.~35, pp.~1909--1926, Nov. 2003.

\bibitem{Johannsen:2013}
T.~Johannsen, ``{PHOTON} {RINGS} {AROUND} {KERR} {AND} {KERR}-{LIKE} {BLACK} {HOLES},'' {\em The Astrophysical Journal}, vol.~777, p.~170, oct 2013.

\bibitem{Grenzebach:2014}
A.~Grenzebach, V.~Perlick, and C.~L\"ammerzahl, ``Photon regions and shadows of kerr-newman-nut black holes with a cosmological constant,'' {\em Phys. Rev.}, vol.~D89, p.~124004, Jun 2014.

\bibitem{Perlick:2017}
V.~Perlick and O.~Y. Tsupko, ``Light propagation in a plasma on kerr spacetime: Separation of the hamilton-jacobi equation and calculation of the shadow,'' {\em Phys. Rev. D}, vol.~95, p.~104003, May 2017.

\bibitem{charbulak_spherical_2018}
D.~Charbulák and Z.~Stuchlík, ``Spherical photon orbits in the field of {Kerr} naked singularities,'' {\em The European Physical Journal C}, vol.~78, p.~879, Nov. 2018.

\bibitem{Johnson_universal_2020}
M.~D. Johnson, A.~Lupsasca, A.~Strominger, G.~N. Wong, S.~Hadar, D.~Kapec, R.~Narayan, A.~Chael, C.~F. Gammie, P.~Galison, D.~C.~M. Palumbo, S.~S. Doeleman, L.~Blackburn, M.~Wielgus, D.~W. Pesce, J.~R. Farah, and J.~M. Moran, ``Universal interferometric signatures of a black hole’s photon ring,'' {\em Science Advances}, vol.~6, p.~eaaz1310, Mar. 2020.

\bibitem{Himwich:2020}
E.~Himwich, M.~D. Johnson, A.~Lupsasca, and A.~Strominger, ``Universal polarimetric signatures of the black hole photon ring,'' {\em Phys. Rev. D}, vol.~101, p.~084020, Apr 2020.

\bibitem{Gelles:2021}
Z.~Gelles, E.~Himwich, M.~D. Johnson, and D.~C.~M. Palumbo, ``Polarized image of equatorial emission in the kerr geometry,'' {\em Phys. Rev. D}, vol.~104, p.~044060, Aug 2021.

\bibitem{Ayzenberg:2022}
D.~Ayzenberg, ``Testing gravity with black hole shadow subrings,'' {\em Classical and Quantum Gravity}, vol.~39, p.~105009, may 2022.

\bibitem{Das:2022}
A.~Das, A.~Saha, and S.~Gangopadhyay, ``Study of circular geodesics and shadow of rotating charged black hole surrounded by perfect fluid dark matter immersed in plasma,'' {\em Classical and Quantum Gravity}, vol.~39, p.~075005, mar 2022.

\bibitem{fathi_spherical_2023}
M.~Fathi, M.~Olivares, and J.~R. Villanueva, ``Spherical photon orbits around a rotating black hole with quintessence and cloud of strings,'' {\em The European Physical Journal Plus}, vol.~138, p.~7, Jan. 2023.

\bibitem{ANJUM2023101195}
A.~Anjum, M.~Afrin, and S.~G. Ghosh, ``Investigating effects of dark matter on photon orbits and black hole shadows,'' {\em Physics of the Dark Universe}, vol.~40, p.~101195, 2023.

\bibitem{Chen:2023}
Y.-X. Chen, J.-H. Huang, and H.~Jiang, ``Radii of spherical photon orbits around kerr-newman black holes,'' {\em Phys. Rev. D}, vol.~107, p.~044066, Feb 2023.

\bibitem{andaru_spherical_2023}
L.~Andaru, A.~Alam, B.~Jayawiguna, and H.~Ramadhan, ``Spherical orbits around {Kerr}-{Newman} and regular black holes,'' preprint, In Review, Oct. 2023.

\bibitem{Gralla:2019}
S.~E. Gralla, D.~E. Holz, and R.~M. Wald, ``Black hole shadows, photon rings, and lensing rings,'' {\em Phys. Rev. D}, vol.~100, p.~024018, Jul 2019.

\bibitem{bisnovatyi-kogan_analytical_2022}
G.~S. Bisnovatyi-Kogan and O.~Y. Tsupko, ``Analytical study of higher-order ring images of the accretion disk around a black hole,'' {\em Physical Review D}, vol.~105, p.~064040, Mar. 2022.

\bibitem{tsupko_shape_2022}
O.~Y. Tsupko, ``Shape of higher-order images of equatorial emission rings around a {Schwarzschild} black hole: {Analytical} description with polar curves,'' {\em Physical Review D}, vol.~106, p.~064033, Sept. 2022.

\bibitem{claudel_geometry_2001}
C.-M. Claudel, K.~S. Virbhadra, and G.~F.~R. Ellis, ``The geometry of photon surfaces,'' {\em Journal of Mathematical Physics}, vol.~42, pp.~818--838, Feb. 2001.

\bibitem{virbhadra_relativistic_2009}
K.~S. Virbhadra, ``Relativistic images of {Schwarzschild} black hole lensing,'' {\em Physical Review D}, vol.~79, p.~083004, Apr. 2009.

\bibitem{virbhadra_compactness_2024}
K.~Virbhadra, ``Compactness of supermassive dark objects at galactic centers,'' {\em Canadian Journal of Physics}, pp.~cjp--2023--0313, June 2024.

\bibitem{Bardeen:1973a}
J.~M. {Bardeen}, W.~H. {Press}, and S.~A. {Teukolsky}, ``{Rotating Black Holes: Locally Nonrotating Frames, Energy Extraction, and Scalar Synchrotron Radiation},'' {\em Astrophys. J}, vol.~178, pp.~347--370, Dec. 1972.

\bibitem{Vazquez:2004}
S.~E. {V{\'a}zquez} and E.~P. {Esteban}, ``{Strong-field gravitational lensing by a Kerr black hole},'' {\em Nuovo Cimento B Serie}, vol.~119, p.~489, May 2004.

\bibitem{Grenzebach:2016}
A.~Grenzebach, {\em The Shadow of Black Holes}, pp.~55--79.
\newblock Cham: Springer International Publishing, 2016.

\bibitem{Hioki:2009na}
K.~Hioki and K.-i. Maeda, ``{Measurement of the Kerr Spin Parameter by Observation of a Compact Object's Shadow},'' {\em Phys. Rev. D}, vol.~80, p.~024042, 2009.

\bibitem{jusufi_constraints_2022}
K.~Jusufi, M.~Azreg-A\"{i}nou, M.~Jamil, and E.~N. Saridakis, ``Constraints on {Barrow} {Entropy} from {M87}* and {S2} {Star} {Observations},'' {\em Universe}, vol.~8, p.~102, Feb. 2022.

\bibitem{okyay_nonlinear_2022}
M.~Okyay and A.~Övgün, ``Nonlinear electrodynamics effects on the black hole shadow, deflection angle, quasinormal modes and greybody factors,'' {\em Journal of Cosmology and Astroparticle Physics}, vol.~2022, p.~009, Jan. 2022.

\bibitem{afrin_parameter_2021}
M.~Afrin, R.~Kumar, and S.~G. Ghosh, ``Parameter estimation of hairy {Kerr} black holes from its shadow and constraints from {M87}*,'' {\em Monthly Notices of the Royal Astronomical Society}, vol.~504, pp.~5927--5940, May 2021.

\bibitem{chen_superradiant_2022}
Y.~Chen, R.~Roy, S.~Vagnozzi, and L.~Visinelli, ``Superradiant evolution of the shadow and photon ring of {Sgr} {A} *,'' {\em Physical Review D}, vol.~106, p.~043021, Aug. 2022.

\bibitem{afrin_tests_2023}
M.~Afrin, S.~Vagnozzi, and S.~G. Ghosh, ``Tests of {Loop} {Quantum} {Gravity} from the {Event} {Horizon} {Telescope} {Results} of {Sgr} {A}*,'' {\em The Astrophysical Journal}, vol.~944, p.~149, Feb. 2023.

\bibitem{sarikulov_shadow_2022}
F.~Sarikulov, F.~Atamurotov, A.~Abdujabbarov, and B.~Ahmedov, ``Shadow of the {Kerr}-like black hole,'' {\em The European Physical Journal C}, vol.~82, p.~771, Sept. 2022.

\bibitem{zahid_shadow_2023}
M.~Zahid, J.~Rayimbaev, F.~Sarikulov, S.~U. Khan, and J.~Ren, ``Shadow of rotating and twisting charged black holes with cloud of strings and quintessence,'' {\em The European Physical Journal C}, vol.~83, p.~855, Sept. 2023.

\bibitem{zubair_4d_2023}
M.~Zubair, M.~A. Raza, F.~Sarikulov, and J.~Rayimbaev, ``{4D} {Einstein}-{Gauss}-{Bonnet} black hole in {Power}-{Yang}-{Mills} field: a shadow study,'' {\em Journal of Cosmology and Astroparticle Physics}, vol.~2023, p.~058, Oct. 2023.

\bibitem{abdujabbarov_coordinate-independent_2015}
A.~A. Abdujabbarov, L.~Rezzolla, and B.~J. Ahmedov, ``A coordinate-independent characterization of a black hole shadow,'' {\em Monthly Notices of the Royal Astronomical Society}, vol.~454, pp.~2423--2435, Dec. 2015.

\bibitem{kumar_black_2020}
R.~Kumar and S.~G. Ghosh, ``Black {Hole} {Parameter} {Estimation} from {Its} {Shadow},'' {\em The Astrophysical Journal}, vol.~892, p.~78, Apr. 2020.

\bibitem{Tsupko:2017a}
O.~Y. Tsupko, ``Analytical calculation of black hole spin using deformation of the shadow,'' {\em Phys. Rev.}, vol.~D95, p.~104058, May 2017.

\bibitem{EventHorizonTelescope:2019dse}
K.~Akiyama {\em et~al.}, ``{First M87 Event Horizon Telescope Results. I. The Shadow of the Supermassive Black Hole},'' {\em Astrophys. J. Lett.}, vol.~875, p.~L1, 2019.

\bibitem{EventHorizonTelescope:2019ggy}
K.~Akiyama {\em et~al.}, ``{First M87 Event Horizon Telescope Results. VI. The Shadow and Mass of the Central Black Hole},'' {\em Astrophys. J. Lett.}, vol.~875, no.~1, p.~L6, 2019.

\bibitem{EventHorizonTelescope:2022xqj}
K.~Akiyama {\em et~al.}, ``{First Sagittarius A* Event Horizon Telescope Results. VI. Testing the Black Hole Metric},'' {\em Astrophys. J. Lett.}, vol.~930, no.~2, p.~L17, 2022.

\bibitem{Tamburini:2019vrf}
F.~Tamburini, B.~Thid\'e, and M.~Della~Valle, ``{Measurement of the spin of the M87 black hole from its observed twisted light},'' {\em Mon. Not. Roy. Astron. Soc.}, vol.~492, no.~1, pp.~L22--L27, 2020.

\bibitem{Daly:2023axh}
R.~A. Daly, M.~Donahue, C.~P. O'Dea, B.~Sebastian, D.~Haggard, and A.~Lu, ``{New black hole spin values for Sagittarius A* obtained with the outflow method},'' {\em Mon. Not. Roy. Astron. Soc.}, vol.~527, no.~1, pp.~428--436, 2023.

\bibitem{EventHorizonTelescope:2022exc}
K.~Akiyama {\em et~al.}, ``{First Sagittarius A* Event Horizon Telescope Results. IV. Variability, Morphology, and Black Hole Mass},'' {\em Astrophys. J. Lett.}, vol.~930, no.~2, p.~L15, 2022.

\bibitem{Hawking:1974rv}
S.~W. Hawking, ``{Black hole explosions},'' {\em Nature}, vol.~248, pp.~30--31, 1974.

\bibitem{belhaj_deflection_2020}
A.~Belhaj, M.~Benali, A.~El~Balali, H.~El~Moumni, and S.-E. Ennadifi, ``Deflection angle and shadow behaviors of quintessential black holes in arbitrary dimensions,'' {\em Classical and Quantum Gravity}, vol.~37, p.~215004, Nov. 2020.

\bibitem{wei_observing_2013}
S.-W. Wei and Y.-X. Liu, ``Observing the shadow of {Einstein}-{Maxwell}-{Dilaton}-{Axion} black hole,'' {\em Journal of Cosmology and Astroparticle Physics}, vol.~2013, pp.~063--063, Nov. 2013.

\bibitem{decanini_fine_2011}
Y.~Décanini, A.~Folacci, and B.~Raffaelli, ``Fine structure of high-energy absorption cross sections for black holes,'' {\em Classical and Quantum Gravity}, vol.~28, p.~175021, Sept. 2011.

\bibitem{li_shadow_2020}
P.-C. Li, M.~Guo, and B.~Chen, ``Shadow of a spinning black hole in an expanding universe,'' {\em Physical Review D}, vol.~101, p.~084041, Apr. 2020.

\bibitem{Nickalls:2006}
R.~W.~D. Nickalls, ``Viète, descartes and the cubic equation,'' {\em The Mathematical Gazette}, vol.~90, no.~518, p.~203–208, 2006.

\bibitem{Zucker:2008}
I.~Zucker, ``92.34 the cubic equation – a new look at the irreducible case,'' {\em The Mathematical Gazette}, vol.~92, no.~524, pp.~264--268, 2008.

\end{thebibliography}

\end{document}